\documentclass[%
 aip,
rsi,%
 amsmath,amssymb,
 reprint,%
]{revtex4-1}

\usepackage[dvipsnames,table]{xcolor}
\usepackage{xcolor}
\usepackage{graphicx}
\usepackage{dcolumn}
\usepackage{bm}
\usepackage{nicefrac}
\newcommand{\n}[1]{\mathbf{#1}}

\begin{document}

\preprint{AIP/123-QED}

\title[Rodr\'{i}guez et al.]{Solving the problem of overdetermination of quasisymmetric equilbrium solutions by near-axis expansions: I. Generalised force balance}

\author{E. Rodr\'{i}guez}
 \altaffiliation[Email: ]{eduardor@princeton.edu}
 \affiliation{ 
Department of Astrophysical Sciences, Princeton University, Princeton, NJ, 08543
}
\affiliation{%
Princeton Plasma Physics Laboratory, Princeton, NJ, 08540
}%

\author{A. Bhattacharjee}
 \altaffiliation[Email: ]{amitava@princeton.edu}
 \affiliation{ 
Department of Astrophysical Sciences, Princeton University, Princeton, NJ, 08543
}
\affiliation{%
Princeton Plasma Physics Laboratory, Princeton, NJ, 08540
}%

\date{\today}

\begin{abstract}
It is well known that the process of construction of quasisymmetric magnetic fields in magnetostatic equilibrium with isotropic pressure suffers from the problem of overdetermination. This has led to the widespread belief that global quasisymmetric solutions are likely not to exist.  We develop a general near-axis expansion procedure that does not rely on the assumption of magnetostatic equilibria with isotropic pressure. We then demonstrate that in equilibria with anisotropic pressure, it is possible to circumvent the problem of overdetermination and carry out the power-series solutions to higher order. This suggests, contrary to current belief, that the existence of globally quasisymmetric fields is likely if one relaxes the assumption of magnetostatic equilibria with isotropic pressure. 
\end{abstract}

\maketitle

\section{Introduction:}\label{sec:intro}

The concept of magnetic confinement as a means to harness fusion energy relies critically on the fact that strong magnetic fields can confine charged particles. To leading order, particles move along magnetic field lines while gyrating around them. However, in the presence of field gradients these particles drift off field lines. While end-losses are eliminated by confining charged particles in a torus, there are restrictions on the types of magnetic field configurations that are efficient in confining particles, even in the absence of collisions. \par
A number of theoretical concepts have been developed that endow magnetic fields with the capability to confine particles. The most restrictive class of strongly confining fields are the so called \textit{isodynamic} configurations. In such configurations, the drift velocities, $v_d$, of all charged particles remain to leading order tangent to magnetic flux surfaces. On the other end of the spectrum we find \textit{omnigeneous} fields. These are a class of fields in which particles, on average, do not have a net drift off magnetic flux surfaces. \par
In between these two extremes lies the concept of \textit{quasisymmetric} (QS) magnetic fields. These are fields in which the dynamics of charged particles are constrained by some approximatedly conserved momenta. A particular example of such a field is one with a continuous symmetry, such as an axisymmetric toroidal plasma. As a result of this continuous symmetry, the axial component of the momentum is exactly conserved. In QS fields momenta are approximately conserved in general, and particles tend to stay close to flux surfaces (generalizing \textit{Tamm's theorem}). The present paper is focused on the construction of QS fields in eqilibrium. \par
In order to construct QS solutions, it has been standard practice in the literature to consider the behaviour of solutions close to the magnetic axis\cite{garrenboozer1991a,landreman2018a,plunk_helander_2018,landreman_sengupta_2019,landreman_sengupta_plunk_2019,jorge_sengupta_landreman_2020}. Garren and Boozer\cite{garrenboozer1991b} consider the construction of magnetostatic (MS) equilibria with isotropic pressure  by expanding the solution around the magnetic axis. They show that in general QS solutions run into the problem of overdetermination when the expansion is carried out to third order in the expansion parameter. (They also show that if one restricts consideration of solutions to those with circular axes, then the overdetermination shows up even earlier, at second order.) While the occurrence of overdetermination cannot, by itself, be regarded as a proof of non-existence of QS magnetostatic equilibria, there has been a tendency in the literature to regard the important finding of Garren and Boozer as strong evidence of the non-existence of global QS solutions. \par
Recent work\cite{burby2019,rodriguez2020} has made it possible to separate the concept of quasisymmetry from all considerations of force balance. This opens up the possibility, undertaken in the present paper, of constructing QS solutions by near-axis expansion without assuming at the outset that the underlying magnetostatic equilibrium has isotropic pressure. This approach turns out to be pivotal in our demonstration that while the problem of overdetermination ails the construction of equilibria with isotropic pressure, the problem is avoided when we consider the plasma pressure to be anisotropic. \par
The following is a plan of this paper, which is the first part of a sequence of two papers. In the following sections we present the relevant equations describing the problem, separating what we will call the \textit{magnetic equations} from the \textit{force-balance} ones. We will then specialize to the case of MS equilbrium with scalar pressure, to recover the results obtained earlier in [\onlinecite{garrenboozer1991b}] . We then present the construction of QS equilibria with anisotropic pressure avoiding the problem of overdetermination. In Part II of the two-paper sequence, we consider the problem of circular axes in some depth and present numerical solutions.

\section{Formulation of Quasisymmetry}
Before dealing with the near-axis expansion, we discuss the mathematical formulation of quasisymmetry in a convenient form. Here we adopt the so called \textit{weak} form of quasisymmetry, as presented in [\onlinecite{rodriguez2020}]. As discussed in Appendix A, this formulation enables an approximate second adiabatic invariant with level surfaces that match flux surfaces. \par
We take the triple product formulation of QS as a starting point,
\begin{equation}
    \nabla\psi\times\nabla B\cdot\nabla(\mathbf{B}\cdot\nabla\mathbf{B})=0, \label{eq:QStripleV}
\end{equation}
where $\psi$ represents the flux surface label, $B$ is the magnitude of the magnetic field and $\mathbf{B}$ is the magnetic field. In using this form, we are assuming implicitly that the magnetic field is not aligned with contours of constant $B$, ensuring the existence of magnetic flux surfaces. Note that equation (1) is derivable from the conditions of quasi-symmetry without making any assumptions regarding force balance\cite{rodriguez2020}.  \par
We now introduce the covariant and contravariant forms of the magnetic field,
\begin{align}
    \mathbf{B}=&\nabla\psi\times\nabla\theta+\iota\nabla\phi\times\nabla\psi=\label{eq:contraB}\\
    =&B_\theta\nabla\theta+B_\phi\nabla\phi+B_\psi\nabla\psi, \label{eq:covarB}
\end{align}
where $\iota$ is a flux function representing the rotational transform of the field, $\{\psi,\theta,\phi\}$ correspond to straight field line coordinates and the covariant functions $B_\theta,~B_\phi,~B_\psi$ are some functions of space. \par
Using the contravariant form (\ref{eq:contraB}) into (\ref{eq:QStripleV}), we obtain
\begin{equation*}
    (\nabla\psi\times\nabla\theta\partial_\theta B+\nabla\psi\times\nabla\phi\partial_\phi B)\cdot\nabla(J^{-1}\partial_\phi B+\iota J^{-1}\partial_\theta B)=0.
\end{equation*}
The function $J$ represents the Jacobian of the straight-field-line coordinate system, taken to be non-zero and well-behaved in the region of interest. Given (\ref{eq:contraB}) and (\ref{eq:covarB}), we can write the Jacobian in the general form
\begin{equation}
    J=(\nabla\psi\times\nabla\theta\cdot\nabla\phi)^{-1}=\frac{B_\phi+\iota B_\theta}{B^2}. \label{eqn:jacCoord}
\end{equation}
At this point, it is convenient to consider a subclass of straight-field-line coordinates, with the Jacobian
$$J=J(\psi,B)=\frac{B_\alpha(\psi)}{B^2},$$
where $B_\alpha\equiv B_\phi+\iota B_\theta$. A special case of such a coordinate system is the Boozer system in the context of magnetohydrostatic equilibrium. It can be shown, however, that for a QS field a coordinate system with such a Jacobian can always be constructed (see Appendix B). We shall refer to such a system as \textit{generalized Boozer coordinates}. Adopting these generalized coordinates, the triple vector product formulation of QS (\ref{eq:QStripleV}) simplifies significantly, and can be rewritten in the form:
\begin{equation}
    (\n{B}\cdot\nabla)\left(\frac{\partial_\phi B}{\partial_\theta B}\right)\partial_\theta B=0.
\end{equation}
This implies that $B=B(\psi,\theta-\Tilde{\alpha}\phi)$ or $B=B(\psi,\phi)$, where $\Tilde{\alpha}=-\partial_\phi B/\partial_\theta B$ is a flux function. To avoid $B=B(\psi)$, we shall take $\tilde{\alpha}$ to be a constant and a rational number. \par
In summary, a field with well-defined flux surfaces is \textit{weakly} quasisymmetric if and only if there is a straight-field-line coordinate system in which the Jacobian has the form $J=B_\alpha(\psi)/B^2$ and the magnetic field magnitude $B=B(\psi,M\theta-N\phi)$ where $N,M\in\mathbb{N}$. It is convenient to define a helical coordinate $\chi=\theta-N\phi/M$. This formulation, in which $B=B(\psi,\chi)$, we call the \textit{Boozer formulation of QS}. \par
Thus, the formulation of quasisymmetry discussed in this section brings the form of the magnetic field very close to the form previously employed in near-axis expansions\cite{landreman2018a,garrenboozer1991b} without making any assumptions regarding force balance.

\section{Expansion Procedure}
Prior to implementing near-axis expansions, we discuss our expansion procedure. We take $\{\psi,\chi,\phi\}$ as our set of independent variables. The magnetic field can then be rewritten as 
\begin{gather}
    \n{B}=B_\theta\nabla\chi+(B_\alpha-\Bar{\iota}B_\theta)\nabla\phi+B_\psi\nabla\psi= \nonumber\\
    =\nabla\psi\times\nabla\chi+\Bar{\iota}\nabla\phi\times\nabla\psi \label{eq:Bcocon}
\end{gather}
where $\Bar{\iota}=\iota-N/M$, leaving the Jacobian unchanged. For the remainder of the paper we shall take $M=1$ as it is a common choice in the literature\cite{garrenboozer1991b,landreman2018a}, excluding the possibility of quasipoloidally symmetric (QPS) arrangements. \par
Having specified the set of independent coordinates, any single-valued space-dependent function $f$ may be written as a Fourier-Taylor series expanded around the magnetic axis, in the form
\begin{equation}
    f(\psi,\theta,\phi)=\sum_{n=0}^\infty\epsilon^n{\sum_{m=0|1}^{n}}\left[f_{nm}^c(\phi)\cos m\chi+f_{nm}^s(\phi)\sin m\chi\right], \label{eq:generalExpansion}
\end{equation}
where the second sum is taken over even or odd indices, depending on the value of $n$.
Here the expansion variable $\epsilon$ is a measure of the distance from the axis, defined as $\epsilon=\sqrt{(\kappa_\mathrm{max})^2\psi/B_\mathrm{min}}$ also a label for flux surfaces around the axis. This is equivalent to the procedure of Garren and Boozer\cite{garrenboozer1991a}, where $B_\mathrm{min}$ is the magnitude of the minimum B-field on axis, and $\kappa_\mathrm{max}$ is the maximum curvature. This choice guarantees that the expansion parameter is dimensionless. The radial dependency of $\epsilon$ imposes a regularity requirement (see [\onlinecite{landreman2018a}]) that forces the $m$-th $\chi$ harmonic to appear, to lowest order, with a power of $\epsilon^m$, which is reflected in the expansion (\ref{eq:generalExpansion}). Finally, for a single-valued $f$, the expansion coefficients $f_{n,m}$ must be periodic in $\phi$. \par
Functions that share the symmetry of the magnetic field have a particularly simple form when written as a power series. For instance, one may write the magnetic field magnitude as
\begin{equation}
    \frac{1}{B^2}=B_0+\sum_{n=1}^\infty\epsilon^n \sum_{m=0|1}^n\left(B_{nm}^c\cos m\chi +B_{nm}^s\sin m\chi\right), \label{eq:Bexp}
\end{equation}
where the expansion coefficients $B_{nm}^{c/s}$ are constant. For functions that depend only on the flux coordinate $\psi$, such as $\iota$ or $B_\alpha$, the Taylor expansion becomes simply,
\begin{equation}
    \iota(\psi)=\sum_{n=0}^\infty\epsilon^{2n}\iota_n,
\end{equation}
where the expansion coefficients are again constants. \par
In order to carry out expansions about the magnetic axis with the chosen set of independent coordinates, we also make use of the inverse map in the same manner as [\onlinecite{garrenboozer1991b}] using the standard dual relations. To complete it, the spatial position vector $\mathbf{x}$ is written in the Frenet basis associated to the magnetic axis. Parametrised by our straight field line coordinates and introducing functions $X,~Y$ and $Z$,\cite{garrenboozer1991a} we write
\begin{align}
    \n{x}=\n{r}_0[l(\phi)]+X(\psi,\chi,\phi)\hat{\n{\kappa}}_0[l(\phi)]+Y(\psi,\chi,\phi)\hat{\n{\tau}}_0+\nonumber\\
    +Z(\psi,\chi,\phi)\hat{\n{b}}_0[l(\phi)].
\end{align}
The vectors correspond to $\n{r}_0$, the magnetic axis (i.e. $\n{x}(\psi=0)=\n{r}_0$), $\hat{\n{\kappa}}_0$, the unit curvature vector, $\hat{\n{\tau}}_0$, the unit binormal and $\hat{\n{b}}_0$, the unit vector tangent to the magnetic axis. The function $l$ is the length along the magnetic axis. Provided that the curvature of the axis is non-vanishing everywhere (shown below), this basis is well-behaved everywhere. A complete description of the orthonormal Frenet basis includes
\begin{gather*}
    \frac{\mathrm{d}\n{r}_0}{\mathrm{d}l}=\n{b}_0 \\
    \frac{\mathrm{d}\hat{\n{b}}_0}{\mathrm{d}l}=\kappa(l)\hat{\kappa}_0 \\
    \frac{\mathrm{d}\hat{\kappa_0}}{\mathrm{d}l}=-\kappa(l)\hat{\n{b}}_0-\tau(l)\hat{\tau}_0 \\
    \frac{\mathrm{d}\hat{\tau}_0}{\mathrm{d}l}=\tau(l)\hat{\kappa}_0.
\end{gather*}
where $\kappa$ and $\tau$ are the curvatures and torsion respectively.

\section{Construction Of The Solution : Magnetic Equations}
Having prescribed the expansion of functions and the appropriate inverse map, we may now construct the fields close to the magnetic axis. Owing to the separation between the considerations of quasisymmetry and those of force-balance, we introduce them at separate stages. The equation for the former will be referred to as the \textit{magnetic equations}, while the latter will be called the \textit{force-balance} equations. The \textit{magnetic equations} consist of two equations: the Jacobian equation and the Co(ntra)variant equation.

\subsection{Jacobian equation}\label{sec:jacEq}
Let us start by focusing on the Jacobian equation, which relates the spatial functions in the inverse map to the magnetic field magnitude. Through (\ref{eq:Bexp}), the Jacobian equation incorporates quasisymmetry explicitly into the construction. \par
There is no unique way of writing this equation. However, we shall capitalise on the form of the Jacobian $J=B_\alpha(\psi)/B^2$ and minimise the amount of algebraic clutter in the equations. Thus, as in [\onlinecite{garrenboozer1991a}], we write
\begin{equation}
    \frac{B_\alpha^2}{B^2}=\left|\frac{\partial\n{x}}{\partial \phi}+\Bar{\iota}\frac{\partial\n{x}}{\partial \chi}\right|^2.
\end{equation} 
Explicitly, 
\begin{align}
    \frac{B_\alpha^2}{B^2}=\left(\Bar{\iota}\partial_\chi X+\partial_\phi X+\tau Y\frac{\mathrm{d}l}{\mathrm{d}\phi}+Z\kappa\frac{\mathrm{d}l}{\mathrm{d}\phi}\right)^2+\nonumber\\
    +\left(\Bar{\iota}\partial_\chi Y+\partial_\phi Y-X\tau\frac{\mathrm{d}l}{\mathrm{d}\phi}\right)^2+\nonumber\\
    +\left(\Bar{\iota}\partial_\chi Z+\partial_\phi Z-X\kappa\frac{\mathrm{d}l}{\mathrm{d}\phi}+\frac{\mathrm{d}l}{\mathrm{d}\phi}\right)^2. \label{eq:Jgen}
\end{align}
In the above form, the equation is almost identical to that in the standard approach\cite{garrenboozer1991a,garrenboozer1991b}, except that here $B_\alpha$ takes the place of the Boozer covariant function $G+\iota I$. \par
In order to expand Eq.(\ref{eq:Jgen}) in a systematic manner, we proceed in two steps. First, we Taylor-expand in $\epsilon$, before explicitly introducing any Fourier series. This allows, in many places, to identify ways in which to simplify the equations that would otherwise be difficult to. Secondly, we substitute the complete Fourier series and collect the different harmonic terms.

\paragraph{Order $\epsilon^0$}:
\begin{equation}
    J^0:~~~~~~B_{\alpha 0}=\frac{\mathrm{d}l}{\mathrm{d}\phi}\frac{1}{\sqrt{B_0}}
\end{equation}
This implies that, as both $B_{\alpha0}$ and $B_0$ are constants by construction, so must $\mathrm{d}l/\mathrm{d}\phi$. Said differently, $l\propto\phi$. This is in fact consistent with the magnetostatic approach\cite{garrenboozer1991b}.

\paragraph{Order $\epsilon^1$}: 
\begin{equation*}
    J^1:~~~~~~-2\left(\frac{\mathrm{d}l}{\mathrm{d}\phi}\right)^2X_1\kappa=B_{\alpha0}^2B_1
\end{equation*}
This may be rewritten explicitly as
\begin{equation*}
    X_1=-\frac{B_1}{2B_0}\frac{1}{\kappa}.
\end{equation*}
The harmonic components can then just be read off to be
\begin{gather}
    X_{11}^C=\frac{\eta}{\kappa} \\
    X_{11}^S=0
\end{gather}
where $\eta=-B_{11}^C/2B_0$. We have chosen by construction $B_{11}^S=0$, which is equivalent to choosing the offset of our angular coordinates and can be done because the magnetic coefficients are constant. \par
From this construction it is clear that, in order to avoid ribbon-like magnetic flux surfaces around the magnetic axis, the axis should have a non-vanishing curvature $\kappa\neq0$ everywhere. Thus we have shown that the observation made in [\onlinecite{garrenboozer1991b}] regarding this requirement on curvature of the magnetic axis does not depend on MS equilbria, but holds for any QS field. This simple picture of flux surface-stretching in the direction of the curvature as the axis is straightened out will become even clearer when we look at the construction of $Y$.

\begin{figure}
    \centering
    \includegraphics[width=0.3\textwidth]{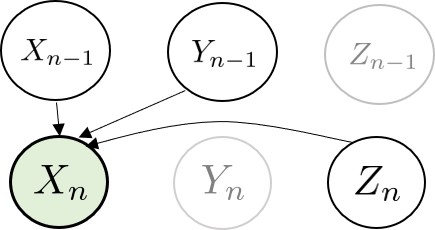}
    \caption{\textbf{Diagram for the Jacobian equation $J^n$.} The green shaded circle represents the function that one should solve the equation for ($X_n$). The arrows point from the leading order of the functions needed to find such a solution.}
    \label{fig:JeqDiag}
\end{figure}

\paragraph{Higher order $\epsilon^n$} After explicitly looking at the first couple of orders (see Appendix C for an explicit construction for $\epsilon^2$), we may extrapolate to arbitrary order $n$. The square of the third parenthesis in (\ref{eq:Jgen}) always yields a term with an isolated $X$ multiplying $\mathrm{d}l/\mathrm{d}\phi$, which makes, to order $\epsilon^n$, solving for $X_n$ simple. Thus, equation $J^n$ can always be used to construct $X_n$ in terms of the functions $Z_n, Y_{n-1}$ and $X_{n-1}$ (and lower orders).This generalisation is represented in Fig. \ref{fig:JeqDiag}. The arrows show the function dependency, while the function shaded in green indicates the function to be solved for. Though it might seem unnecessary in this simple case, such representations will prove to be useful later. \par
In short, the $J^n$ equation contributes $n+1$ independent constraints, uniquely defining the form of $X_n$ at every order.

\subsection{Co(ntra)variant equation}
The \textit{co(ntra)variant} equation refers to the vector equation relating the covariant and contravariant forms of the magnetic field as represented by (\ref{eq:Bcocon}). This equation includes information about the existence of flux surfaces and the divergenceless nature of the magnetic field. It is straightforward to cast it in the form
\begin{align}
    (B_\alpha-&\Bar{\iota}B_\theta)\frac{\partial\n{x}}{\partial\psi}\times\frac{\partial\n{x}}{\partial\chi}+B_\theta\frac{\partial\n{x}}{\partial\phi}\times\frac{\partial\n{x}}{\partial\psi}+B_\psi\frac{\partial\n{x}}{\partial\chi}\times\frac{\partial\n{x}}{\partial\phi}=\frac{\partial\n{x}}{\partial\phi}+\nonumber\\
    &+\Bar{\iota}(\psi)\frac{\partial\n{x}}{\partial\chi}. \label{eq:co(ntra)variant}
\end{align}
 \par
The following are its three Frenet components: the $\hat{\n{b}}_0$ component,
\begin{align}
    &-(B_\alpha-\bar{\iota}B_\theta)\left(\partial_\chi X\partial_\psi Y-\partial_\psi X\partial_\chi Y\right)-\nonumber\\
    &-B_\psi\left[\partial_\chi Y\left(\partial_\phi X+\tau Y\frac{\mathrm{d}l}{\mathrm{d}\phi}+Z\kappa\frac{\mathrm{d}l}{\mathrm{d}\phi}\right)-\partial_\chi X\left(\partial_\phi Y-\right.\right.\nonumber\\
    &\left.\left.-X\tau\frac{\mathrm{d}l}{\mathrm{d}\phi}\right)\right]+B_\theta\left[\partial_\psi Y\left(\partial_\phi X+\tau Y\frac{\mathrm{d}l}{\mathrm{d}\phi}+Z\kappa\frac{\mathrm{d}l}{\mathrm{d}\phi}\right)-\right.\nonumber\\
    &\left.-\partial_\psi X\left(\partial_\phi Y-X\tau\frac{\mathrm{d}l}{\mathrm{d}\phi}\right)\right]=\nonumber\\
    &=\left(\partial_\phi Z-X\kappa\frac{\mathrm{d}l}{\mathrm{d}\phi}+\frac{\mathrm{d}l}{\mathrm{d}\phi}\right)+\Bar{\iota}\partial_\chi Z, \label{eq:Cb}
\end{align}
the $\hat{\kappa}_0$ component,
\begin{align}
    &-(B_\alpha-\bar{\iota}B_\theta)\left(\partial_\chi Y\partial_\psi Z-\partial_\psi Y\partial_\chi Z\right)-\nonumber\\
    &-B_\psi\left[\partial_\chi Z\left(\partial_\phi Y-X\tau\frac{\mathrm{d}l}{\mathrm{d}\phi}\right)-\partial_\chi Y\left(\partial_\phi Z-X\kappa\frac{\mathrm{d}l}{\mathrm{d}\phi}+\frac{\mathrm{d}l}{\mathrm{d}\phi}\right)\right]+\nonumber\\
    &+B_\theta\left[\partial_\psi Z\left(\partial_\phi Y-X\tau\frac{\mathrm{d}l}{\mathrm{d}\phi}\right)-\partial_\psi Y\left(\partial_\phi Z-X\kappa\frac{\mathrm{d}l}{\mathrm{d}\phi}+\frac{\mathrm{d}l}{\mathrm{d}\phi}\right)\right]=\nonumber\\
    &=\left(\partial_\phi X+\tau Y\frac{\mathrm{d}l}{\mathrm{d}\phi}+Z\kappa\frac{\mathrm{d}l}{\mathrm{d}\phi}\right)+\Bar{\iota}\partial_\chi X, \label{eq:Ck}
\end{align}
and the $\hat{\tau}_0$ component,
\begin{align}
    &-(B_\alpha-\bar{\iota}B_\theta)\left(\partial_\chi Z\partial_\psi X-\partial_\psi Z\partial_\chi X\right)-\nonumber\\
    &-B_\psi\left[\partial_\chi X\left(\partial_\phi Z-X\kappa\frac{\mathrm{d}l}{\mathrm{d}\phi}+\frac{\mathrm{d}l}{\mathrm{d}\phi}\right)-\partial_\chi Z\left(\partial_\phi X+\tau Y\frac{\mathrm{d}l}{\mathrm{d}\phi}+\right.\right.\nonumber\\
    &\left.\left.+Z\kappa\frac{\mathrm{d}l}{\mathrm{d}\phi}\right)\right]+B_\theta\left[\partial_\psi X\left(\partial_\phi Z-X\kappa\frac{\mathrm{d}l}{\mathrm{d}\phi}+\frac{\mathrm{d}l}{\mathrm{d}\phi}\right)-\right.\nonumber\\
    &\left.-\partial_\psi Z\left(\partial_\phi X+\tau Y\frac{\mathrm{d}l}{\mathrm{d}\phi}+Z\kappa\frac{\mathrm{d}l}{\mathrm{d}\phi}\right)\right]=\nonumber\\
    &=\left(\partial_\phi Y-X\tau\frac{\mathrm{d}l}{\mathrm{d}\phi}\right)+\Bar{\iota}\partial_\chi Y. \label{eq:Ct}
\end{align}
These equations will be referred to in shorthand as $C_b$ and $C_\perp$ (the latter including both (\ref{eq:Ck}) and (\ref{eq:Ct}), as they usually appear together). Once expanded, these equations closely resemble those in [\onlinecite{garrenboozer1991b}], except for the main difference that $B_\theta$ is not necessarily a flux function in our case.

\paragraph{Order $\epsilon^{-1}$}: the $C_\perp$ equations have a leading $\epsilon^{-1}$ order, due to the presence of flux derivatives. We write
\begin{gather*}
    C_\kappa^{-1}:~~~~-B_{\theta0}\frac{\mathrm{d}l}{\mathrm{d}\phi}Y_1=0 \\
    C_\tau^{-1}:~~~~-B_{\theta0}\frac{\mathrm{d}l}{\mathrm{d}\phi}X_1=0.
\end{gather*}
Both these equations are satisfied if we take
\begin{equation}
    B_{\theta0}=0.
\end{equation}
We assume that neither $X_1$, $Y_1$ nor $\mathrm{d}l/\mathrm{d}\phi$ vanish trivially, based on considerations of regularity at the magnetic axis.

\paragraph{Order $\epsilon^0$}:
\begin{gather*}
    C_b^0:~~~~-2\frac{\mathrm{d}l}{\mathrm{d}\phi}+B_{\alpha0}\left(X_1\partial_\chi Y_1-Y_1\partial_\chi X_1\right)=0 \\
    C_\kappa^0:~~~~-B_{\theta1}\frac{\mathrm{d}l}{\mathrm{d}\phi} Y_1+B_{\alpha0}\left(Y_1\partial_\chi Z_1-Z_1\partial_\chi Y_1\right)=0 \\
    C_\tau^0:~~~~B_{\theta1}\frac{\mathrm{d}l}{\mathrm{d}\phi} X_1+B_{\alpha0}\left(Z_1\partial_\chi X_1-X_1\partial_\chi Z_1\right)=0 
\end{gather*}
Combining the latter two conditions above to eliminate $B_{\theta1}$, and applying $C_b^0$, we find,
\begin{equation}
    Z_1\frac{\mathrm{d}l}{\mathrm{d}\phi}=0\rightarrow Z_1=0,
\end{equation}
and therefore also,
\begin{equation}
    B_{\theta1}=0.
\end{equation}
The only other remaining equation is $C_b^0$. Using the expansions for $X$ and $Y$, we obtain

\begin{equation}
    Y_{11}^S=\frac{2\sqrt{B_0}}{\eta}\kappa.
\end{equation}
Thus, as the curvature on axis becomes smaller, we expect the flux surfaces close to the magnetic axis to get squeezed in the direction of the binormal $\hat{\tau}$, and the surface becomes more elongated along the direction of the curvature.

\paragraph{Order $\epsilon^1$}: let us start by focusing on the $C_\perp^1$ equations, leaving aside the $\hat{b}_0$ component for later. We may write,
\begin{align*}
    C_\kappa^1:~~~~~-B_{\theta2}\frac{\mathrm{d}l}{\mathrm{d}\phi}Y_1+2B_{\psi0}\frac{\mathrm{d}l}{\mathrm{d}\phi}\partial_\chi Y_1-B_{\alpha0}(2Z_2\partial_\chi Y_1-\nonumber\\
    -Y_1\partial_\chi Z_2)=2\left(\frac{\mathrm{d}l}{\mathrm{d}\phi}\tau Y_1+\partial_\phi X_1+\Bar{\iota}_0\partial_\chi X_1\right)
\end{align*}
\begin{align*}
    C_\tau^1:~~~~~B_{\theta2}\frac{\mathrm{d}l}{\mathrm{d}\phi}X_1-2B_{\psi0}\frac{\mathrm{d}l}{\mathrm{d}\phi}\partial_\chi X_1+B_{\alpha0}(2Z_2\partial_\chi X_1-\nonumber\\
    -X_1\partial_\chi Z_2)=2\left(-\frac{\mathrm{d}l}{\mathrm{d}\phi}\tau X_1+\partial_\phi Y_1+\Bar{\iota}_0\partial_\chi Y_1\right).
\end{align*}
The largest order functions in these equations are $Z_2$ and $B_{\theta2}$, so it is natural to solve for these.\par
Combining $C_\kappa^1$ and $C_\tau^1$, and applying our knowledge from $C_b^0$, we obtain
\begin{equation}
    Z_2=\sqrt{B_0}B_{\psi0}-\left(\frac{\mathrm{d}l}{\mathrm{d}\phi}\right)^{-1}(\partial_\phi+\Bar{\iota}_0\partial_\chi)\left(\frac{X_1^2+Y_1^2}{4}\right).
\end{equation}
Note here that $Z_2$ depends on $B_{\psi0}$, generally a function of toroidal angle $\phi$ which will not be constrained until some form of force balance is assumed. \par
From this form, the harmonic components of $Z_2$ may be easily obtained (using the notation from [\onlinecite{garrenboozer1991a}]), 
\begin{gather*}
    Z_{2,0}=\frac{\mathrm{d}l}{\mathrm{d}\phi}\frac{B_{\psi0}}{B_{\alpha0}}-\frac{1}{8}\left(\frac{\mathrm{d}l}{\mathrm{d}\phi}\right)^{-1}\frac{\mathrm{d}V_1}{\mathrm{d}\phi} \\
    Z_{2,2}^C=-\frac{1}{8}\left(\frac{\mathrm{d}l}{\mathrm{d}\phi}\right)^{-1}\left[\frac{\mathrm{d}V_3}{\mathrm{d}\phi}+2\Bar{\iota}_0V_2\right] \\
    Z_{2,2}^S=-\frac{1}{8}\left(\frac{\mathrm{d}l}{\mathrm{d}\phi}\right)^{-1}\left[\frac{\mathrm{d}l}{\mathrm{d}\phi}-2\Bar{\iota}_0V_3\right]
\end{gather*}
where,
\begin{gather*}
    V_1 = (X_{1,1}^C)^2+(Y_{1,1}^C)^2+(Y_{1,1}^S)^2 \\
    V_2 = 2Y_{1,1}^CY_{1,1}^S \\
    V_3 = (X_{1,1}^C)^2+(Y_{1,1}^C)^2-(Y_{1,1}^S)^2.
\end{gather*}
To arrive at this form of $Z_2$, we have made use of three constraint equations. To see what the remaining equations are, we may substitute the freshly obtained forms for $Z_2$ into either equation $C_\kappa^1$ or $C_\tau^1$.
Two of the components, those corresponding to the largest harmonics, read
\begin{gather*}
    0=B_{\theta22}^SX_{11}^C+B_{\theta22}^CX_{11}^S \\
    0=B_{\theta22}^CX_{11}^C-B_{\theta22}^SX_{11}^S,
\end{gather*}
from which it follows that
\begin{equation}
    B_{\theta22}=0.
\end{equation}
The last remaining constraint may be written as an ordinary differential equation (ODE) for the function $Y_{11}^C$,
 \begin{align*}
    (Y_{11}^C)'-Y_{11}^C\frac{\kappa'}{\kappa}+&(Y_{11}^C)^2\frac{\Bar{\iota}_0\eta}{2\sqrt{B_0}\kappa}+\Bar{\iota}_0\left(\frac{2\sqrt{B_0}}{\eta}\kappa+\frac{\eta^3}{2\sqrt{B_0}}\frac{1}{\kappa^3}\right)-\\
    &-\frac{\mathrm{d}l}{\mathrm{d}\phi}(2\tau+B_{\theta20})\frac{\eta}{\kappa}=0.
 \end{align*}
The interpretation of this equation as an ODE and not an algebraic equation for the function $B_{\theta 20}$ will be clear when we have the opportunity to look into the force balance equations, which may be written in a form independent of the spatial functions $X,~Y$ and $Z$. Defining $Y_{11}^C=Y_{11}^S\sigma$, the equation above reduces to,
 \begin{equation}
     \frac{\mathrm{d}\sigma}{\mathrm{d}\phi}=-\Bar{\iota}_0\left[1+\frac{1}{4B_0}\left(\frac{\eta}{\kappa}\right)^4+\sigma^2\right]+\frac{B_{\alpha0}}{2}(2\tau+B_{\theta20})\left(\frac{\eta}{\kappa}\right)^2. \label{eq:ODEY11C}
 \end{equation}
This first order non-linear differential equation is to be solved for $\sigma$ subject to periodic boundary conditions $\sigma(0)=\sigma(2\pi)$. Such an equation has been previously analysed thoroughly\cite{landreman2018a,garrenboozer1991b}, with the only difference that in our generalised case $B_{\theta20}$ is not a constant but an unknown function of $\phi$.  \par
Summarising, the following are results obtained from considerations of $C_\perp^1$: we have constructed explicitly $Z_2$ in terms of $Y_{11}^C$, $B_{\psi0}$ and other known functions, obtained a differential equation for $Y_{11}^C$ and pinned down two of the components of $B_{\theta2}$. \par
Let us now turn our attention to $C_b^1$, and write
\begin{align*}
   C_b^1:~~~~ 2\frac{\mathrm{d}l}{\mathrm{d}\phi}X_1\kappa=B_{\alpha0}(2Y_2\partial_\chi X_1+Y_1\partial_\chi X_2-\\
   -2X_2\partial_\chi Y_1-X_1\partial_\chi Y_2).
\end{align*}
From the harmonic components of the equation, we obtain
\begin{gather}
    Y_{22}^C=Y_{20}-\frac{1}{B_{\alpha0}}\frac{\beta}{X_{11}^C}, \\
    Y_{22}^S=-\frac{1}{B_{\alpha0}}\frac{\alpha}{X_{11}^C},
\end{gather}
where,
\begin{gather*}
    \alpha=\frac{\mathrm{d}l}{\mathrm{d}\phi}\kappa X_{11}^C-B_{\alpha0}( X_{22}^S Y_{11}^C- X_{20}^C Y_{11}^S- X_{22}^C Y_{11}^S) \\
    \beta=-B_{\alpha0}( X_{22}^C Y_{11}^C- X_{20}^C Y_{11}^C+ X_{22}^S Y_{11}^S).
\end{gather*} 
These two equations from $C_b^1$ prescribe the form of the second harmonics of $Y_2$, leaving $Y_{20}$ unconstrained.

\paragraph{Higher order $\epsilon^n$}: Following the steps taken for the lower order expansions, we can obtain an explicit closed form for $Z_{n+1}$ from $C_\perp^n$. It is important to note that the $Z$ function of interest is an order higher  than the equation, which thus includes terms in $B_{\theta n}$, $B_{\psi n-1}$, $Z_n$, $Y_n$ and $X_n$ (and lower orders). These constitute $n+2$ constraint equations; for more details, we refer to Appendix D. \par

\begin{figure}
    \centering
    \includegraphics[width=0.4\textwidth]{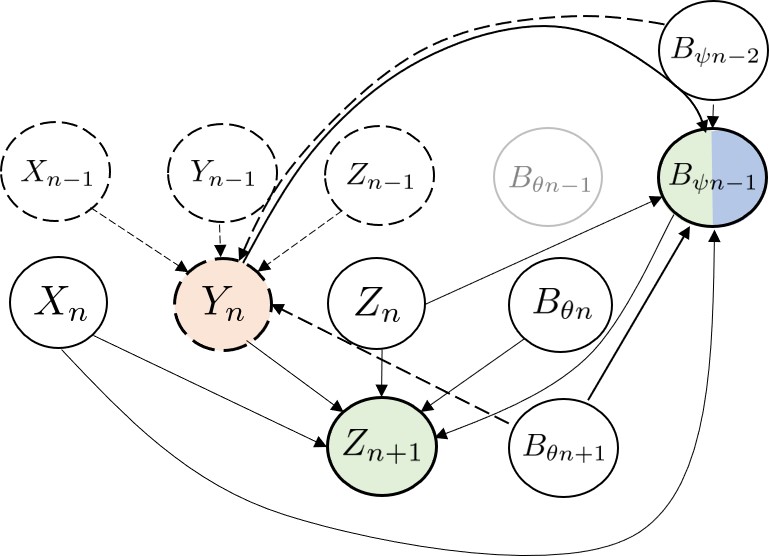}
    \caption{\textbf{Diagram for the $C_\perp^n$ equations.} The green shaded circles represent the functions that one should solve the equation for ($B_{\psi n-1},~Z_{n+1}$). In this case, $Y_n$ has been shaded reddish to indicate that every other order, one should in fact solve for one of the components of $Y$ as well. The blue shade represents that the $B_{\psi n0}$ functions are not pinned down by the equations. The arrows indicate the leading order of the functions needed to find a solution (with broken lines representing only even order case).}
    \label{fig:CperpeqDiag}
\end{figure}

In much the same way as it occurred at lower order, the largest harmonics of $B_{\theta n}$ must also vanish to arbitrary high order $n$. With $B_{\theta n+1,n+1}$ appearing in a trivial way, this takes up, to order $\epsilon^n$, the place of four constraint equations (corresponidng to the largest harmonic of both $C_\kappa$ and $C_\tau$). This leaves a remaining total of $n$ constraints. The $n$ remaining equations are used to construct solutions for the function $B_\psi$. An explicit general construction can be found in Appendix D. Here we just point that $B_{\psi n-1}$ can be found in terms of $B_{\theta n+1},~B_{\psi n-2},~X_{n},~Y_{n}$ and $Z_{n}$. This is not quite true for $B_{\psi n 0}$, a case that deserves special consideration. As a straightforward extension of order $\epsilon^1$ and Eq.(\ref{eq:ODEY11C}), it is convenient to interpret the equation for $B_{\psi n0}$ as a differential equation for $Y_{n}$. This occurs at every other order (when the order of expansion is even), and leaves $B_{\psi n 0}$ as a free function. The ODE also depends on $B_{\theta n0}$, but no other component of $B_\theta$ directly. We represent these $C_\perp$ equations diagramatically in Fig. \ref{fig:CperpeqDiag}.
\par
\begin{figure}
    \centering
    \includegraphics[width=0.45\textwidth]{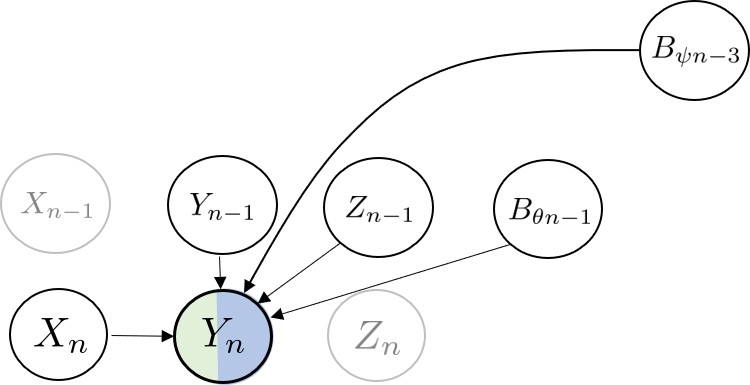}
    \caption{\textbf{Diagram for the $C_b^n$ equation.} The green shaded circle represents the function that one should solve the equation for ($Y_n$). The blue shade represents the fact that the $Y_{n0}$ functions are not pinned down by the equations. The arrows indicate the leading order of the functions needed to find a solution.}
    \label{fig:CbeqDiag}
\end{figure}

Concerning the $\hat{\n{b}}_0$ component of the co(ntra)variant equation, it has $n+1$ independent constraint equations. These will be used to solve for $Y_{n+1}$ in terms of functions $B_{\psi n-2},~X_{n+1},~B_{\theta n}$ and $Z_n$. The constraint equations are enough to pin down $n+1$ of the $n+2$ components of $Y_{n+1}$. It is worth remarking on the remaining unconstrained functions. As discussed above, one of the $C_\perp$ constraints should be employed as an ODE for $Y$. The remaining $Y$ degree of freedom is $Y_{n0}$. However, it remains unconstrained at the level of the magnetic equations. We again summarise these findings in a diagram (see Fig. \ref{fig:CbeqDiag}).
\par
In summary, and re-counting the functions and constraint equations at our disposal, we present Table \ref{tab:countingQSJCaxis}. The magnetic equations are enough to determine $X$ and $Z$, as well as $B_\psi$ and $Y$ up to a flux function, constraining some of the components of $B_\theta$, which is otherwise free. At every other order, we need to solve an ODE. 

\begin{table}[]
    \centering \hspace*{-.5cm}
    \begin{tabular}{|c||c|c|c|}
        \hline
    Eqn. & Order & Solve for\dots & Nb. eqns \\ \hline\hline
    $J^n$ & $n$ & $X_n$ & $n+1$ \\\hline
    $C_b^n$ & $n$ & $Y_{n+1}$ & $n+1$ \\\hline 
    $C_\perp^n$ & $n=2k$ & $B_{\psi n-1}$ & $n$ \\
     &  & $Z_{n+1}$ & $n+2$ \\
     &  & $B_{\theta n+1,n+1}$ & $2*$ \\ \hline
    $C_\perp^n$ & $n=2k+1$ & $Y_n$ & $1"$ \\
     &   & $B_{\psi n-1}$ & $n-1$ \\
     &   & $Z_{n+1}$ & $n+2$ \\
     &   & $B_{\theta n+1,n+1}$ & $2*$ \\\hline
    \end{tabular}
    \caption{\textbf{Counting of equations and degrees of freedom for magnetic equations.} each column shows: the label of equations, the order of expansion, what the equations are solved for and the number of consraints (or independent equations) they amount to. The asterisk indicates that due to a trivial solution, two other consteraints are also satisfied by the trivial solution (but we drop them so that the total counting can be done correctly). The " indicates that a differential equation needs to be solved subject to periodic boundary conditions, unlike the rest of equations which simply require algebraic manipulations.}
    \label{tab:countingQSJCaxis}
\end{table}

\section{Case of isotropic pressure}

In this section we apply our general procedure to the well-known case of MS equilibrium. While our results in this case are no different than those of [\onlinecite{garrenboozer1991b}], it is not only important as a check of the correctness of our approach, but also as a step in how to best include the force-balance condition in the construction. So we begin with
\begin{equation*}
    \n{j}\times\n{B}=\nabla p,
\end{equation*}
where $p$ represents a scalar pressure. We shall consider $p$ to be a general function of space with no particular symmetry.

\subsection{Constructing force-balance equations}
To efficiently bring the magnetic construction of the previous section into contact with the MS force balance condition, we need to write the latter in an appropriate form. Making extensive use of the covariant and contravariant forms of the magnetic field, the left-hand-side (LHS) of the force-balance equation may be written as,
\begin{align}
    \n{j}&\times\n{B}=\nonumber\\
    &-J^{-1}\left\{\Bar{\iota}\left(\partial_\psi B_\theta-\partial_\chi B_\psi\right)-\left[\partial_\phi B_\psi-\partial_\psi B_\alpha +\right.\right.\nonumber\\
    &\left.\left.+\partial_\psi(\Bar{\iota}B_\theta)\right]\right\}\nabla\psi-J^{-1}\left(\partial_\chi B_\alpha-\Bar{\iota}\partial_\chi B_\theta-\right.\nonumber\\
    &\left.-\partial_\phi B_\theta\right)(\Bar{\iota}\nabla\phi-\nabla\chi), \label{eq:lhsFBal}
\end{align}
where $J$ is again the Jacobian of our generalized Boozer coordinate system. Given the form of (\ref{eq:lhsFBal}), the right-hand-side (RHS) of the force-balance equation is naturally written as,
\begin{equation}
    \nabla p=\partial_\psi p\nabla\psi+\partial_\chi p \nabla\chi+\partial_\phi p\nabla\phi.\label{eq:rhsFB}
\end{equation}
From the vector equation, three scalar equations may be read out by projecting along the dual of $\{\nabla\psi,\nabla\chi,\nabla\phi\}$ which by construction has a non-zero Jacobian everywhere. This gives
\begin{gather}
    \mathrm{III}:~~~~\Bar{\iota}\left(\partial_\psi B_\theta-\partial_\chi B_\psi\right)-\left[\partial_\phi B_\psi-\partial_\psi B_\alpha +\partial_\psi(\Bar{\iota}B_\theta)\right]+\nonumber\\
    +J\partial_\psi p=0~~~~~~~~~~~~ \label{eq:MSIII} \\
    \mathrm{II}:~~~~~ \Bar{\iota}\partial_\chi B_\theta+\partial_\phi B_\theta-J\partial_\chi p=0 \label{eq:MSII}\\
    -\Bar{\iota}\left(\Bar{\iota}\partial_\chi B_\theta+\partial_\phi B_\theta\right)+J\partial_\phi p=0. \nonumber
\end{gather}
The second equation may be combined with the latter to yield,
\begin{equation}
    \mathrm{I}:~~~~~~~~(\partial_\phi+\Bar{\iota}\partial_\chi)p=0. \label{eq:b.gP0}
\end{equation}

\subsection{Expansion procedure: magnetostatic force balance}

\subsubsection{Equation I}
Equation I is a coordinate representation of the magnetic equation $\n{B}\cdot\nabla p=0$, and thus we expect the solution to the equation to be that the scalar pressure $p=p(\psi)$. Let us see how this shows up order by order.
\paragraph{Order $\epsilon^0$:} 
\begin{equation*}
    \partial_\phi p_0=0 \rightarrow p_0=\mathrm{const.}
\end{equation*}

\paragraph{Order $\epsilon^1$:}
\begin{gather*}
    p_{11}^C{}'+\Bar{\iota}_0p_{11}^S=0 \\
    p_{11}^S{}'-\Bar{\iota}_0p_{11}^C=0
\end{gather*}
Substitute one into the other to obtain,
\begin{equation*}
    \left(p_{11}^C\right)''+\Bar{\iota}_0^2 p_{11}^C=0.
\end{equation*}
which is an equation analogus to a simple harmonic oscillator (SHO) with a solution of the form $p_{11}^C\sim\exp(\pm i\Bar{\iota}_0\phi)$. For a rotational transform that is generally irrational, this solution is not periodic in $\phi$ (meaning that $p_{11}^C(0)\neq p_{11}^C(2\pi)$). As $p$ is a physically meaningful quantity, its coefficients must be periodic in the angular coordinates, so
\begin{equation*}
    p_{11}^C=0=p_{11}^S
\end{equation*}

\paragraph{Higher order:} an identical argument to that for $O(\epsilon^1)$ holds for arbitrarily large order, forcing the pressure to be a flux function, 
\begin{equation}
    p=p(\psi).
\end{equation}

\subsubsection{Equation II}
From (\ref{eq:MSII}), and using the relation $p=p(\psi)$, we obtain
\begin{equation*}
    \Bar{\iota}\partial_\chi B_\theta+\partial_\phi B_\theta=0.
\end{equation*}
Hence, $B_\theta=B_\theta(\psi)$.

\subsubsection{Equation III}

\paragraph{Order $\epsilon^0$}: 
\begin{equation*}
    -B_{\psi0}'+B_{\alpha1}+B_{\alpha0}B_0 p_{20}=0
\end{equation*}
from which it follows that $B_{\psi0}$ is a constant and we obtain an additional condition relating constant coefficients,
\begin{equation*}
    B_0p_{20}=-\frac{B_{\alpha1}}{B_{\alpha0}}.
\end{equation*}

\paragraph{Order $\epsilon^1$}: 
\begin{gather*}
    \Bar{\iota}_0B_{\psi11}^S-B_{\alpha0}B_{11}^C p_{20}+B_{\psi11}^C{}'=0 \\
    \Bar{\iota}_0 B_{\psi11}^C=B_{\psi11}^S{}'
\end{gather*}
These may be combined into SHO equations for $B_\psi$. The requirement on periodicity allows only for a trivial general solution, leaving the particular solution. Thus,
\begin{gather*}
    B_{\psi11}^S=\frac{B_{\alpha0}}{\Bar{\iota}_0}B_{11}^C p_{20} \\
    B_{\psi11}^C=0.
\end{gather*} 
So the coefficients of $B_\psi$ end up being constant. It follows that to this order, $B_\psi$ shares the same angular dependence as $1/B^2$.

\paragraph{Higher order $\epsilon^n$}: a general expression can be obtained by looking at higher orders of the expansion. However, it suffices to note that the set defines the function $B_\psi$ to the $n$-th order, and that it has the same symmetry as the magnitude of the quantity $1/B^2$. In total, these constitute $n+1$ constraints to order $\epsilon^n$.\par 
The three sets of constraints from the MS force balance equations are summarised in Table \ref{tab:countingQSMSaxis}.

\begin{table}[]
    \centering \hspace*{-.5cm}
    \begin{tabular}{|c||c|c|c|}
        \hline
    Eqn. & Order & Solve for\dots & Nb. eqns \\ \hline\hline
    I & $n$ & $p_n$ & $n+1$ \\\hline
    II & $n$ & $B_{\theta n}$ & $n+1$ \\\hline 
    III & $n$ & $B_{\psi n}$ & $n+1$ \\\hline
    \end{tabular}
    \caption{\textbf{Counting of equations and degrees of freedom.} Each column shows: the label of the equations, the order of expansion, what the equations are solved for and the number of consraints (or independent equations) they amount to. Note that in the magnetostatic case the dependency of the equations mostly results in constants.}
    \label{tab:countingQSMSaxis}
\end{table}

\subsection{Complete near-axis construction for isotropic pressure }
With both the magnetic and MS force-balance parts of the near-axis expansion in place, we now bring these parts together. Towards that end, we shall make use of the information in Tables \ref{tab:countingQSJCaxis} and \ref{tab:countingQSMSaxis}, which are essential to complete an account of equations and functions and implement the construction. \par
\begin{figure}
    \includegraphics[width=0.5\textwidth]{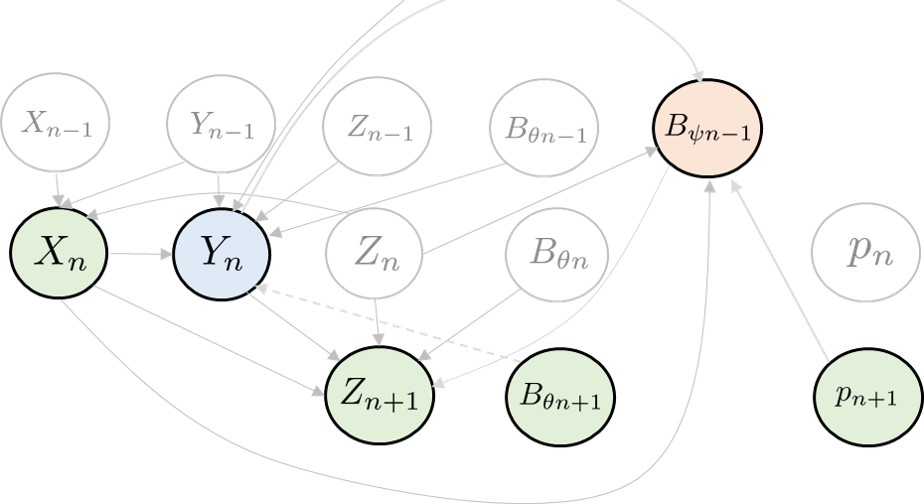}
    \caption{\textbf{Consistent order of MS functions.} Order of functions that ought to be simultaneously solved for. The color code represents: green - precise constraint number for the number of functions, blue - additional free functions every even order, orange - additional constraints.}
    \label{fig:ConstructNAMS}
\end{figure}

First, it is important to know, for a given order, how the different functions come into play. The question that needs to be addressed is: how can we construct higher orders of the expansion in such a way that the problem formulated is self-consistent? The answer is shown schematically in Fig. \ref{fig:ConstructNAMS}. \par
Second, we should indicate precisely which equations are needed to solve for the various functions. This collection is shown in Table \ref{tab:countingQSMSaxis}. \par
\begin{table}
    \includegraphics[width=0.45\textwidth]{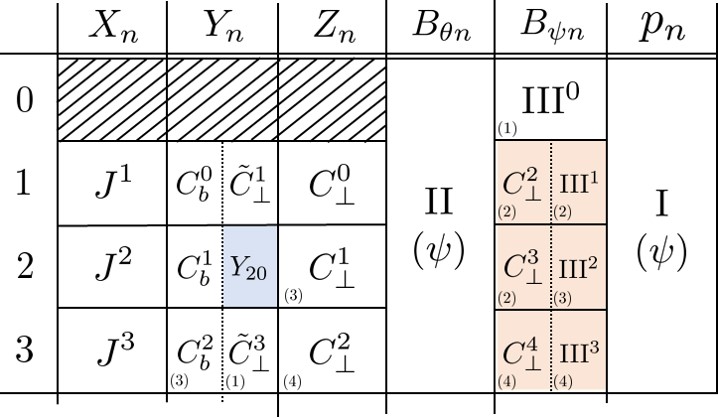}
    \caption{\textbf{Magnetostatic near-axis construction.} Each cell indicates the equations that need to be used to find a function (column) at a given order (row). The equation labels are the ones introduced in the text: $J$ for the Jacobian equation, $C_\perp$ and $C_b$ for the co(ntra)variant perpendicular and $b_0$ equations, and I, II and III for the force balance equations. The numbers in parenthesis denote the number of constraint equations (where no number is given is because they account for the precise number of unknowns in that cell or is unnecessary). The undivided 4th and last columns represent the fact that the functions are flux functions. The blue color shows that a new free function is being introduced, while the light orange represents overdetermination (multiple equations for the same function). }
    \label{tab:eqConstructNAMS}
\end{table}
With these two tools in hand, we are now in a position to determine, by counting equations and functions whether the construction is overdetermined or not. As concluded in [\onlinecite{garrenboozer1991b}], it \textit{is} evident that the problem is overconstrained. The heart of the problem lies in the function $B_\psi$, a function that is both determined by the magnetic as well as the force-balance equations (highlighted orange in Tab. \ref{tab:eqConstructNAMS}). Counting the number of constraints, we see that at a given order $n$, we have $n$ or $n-1$ (depending on whether the order is even or odd, respectively) additional constraints that need to be satisfied. \par
The first occurrence of this equation surplus appears at second order ($n=2$). Combining the magnetic equations describing $B_{\psi11}$ together with the force balance equation III, one obtains two coupled differential equations that are to be solved for $Y_{20}$ and a characteristic function of the magnetic axis ($\kappa$ for instance). No other free function remains in the problem. \par
If the construction is to be successful to arbitrary order, there should be additional free functions appearing at every order in a number at least equal to the number of extra constraints that appear at each order. This is not the case, as  $Y_{n0}$ is introduced only at every other order. Thus, $n-1$ over-constraining equations lack free functions to be solved for beyond the second order. \par
Thus, in the MS limit, we arrive at the same conclusions as previously shown in [\onlinecite{garrenboozer1991b}]. This, however, does not settle one way or another the question of existence of MS equilibria because over-determined equations can still have solutions.

\section{Case of anisotropic pressure} \label{NAEMS}

We saw in Section V how the MS equilibrium solutions become overdetermined close to the magnetic axis to third order. However,  formally speaking, there is no reason why one should stick to MS equilibrium. A natural hypothesis is that by introducing a more general form of force balance, which might include additional degrees of freedom, the overdetermination problem can be solved. We remark that there does not seem to be any \textit{fundamental} breakdown of the expansion procedure for MS equilbria other than there being too little freedom. \par
In this section we consider the case of anisotropic pressure as an example of a force balance with more freedom. Including a pressure tensor is the most immediate and natural extension to MS equilibria. Let us consider the force balance given by
\begin{equation}
    \n{j}\times\n{B}=\nabla\cdot\Pi,
\end{equation}
where,
\begin{equation}
    \Pi=(p_\parallel-p_\perp)\n{b}\n{b}+p_\perp \mathbb{I}.
\end{equation}
The unit dyad is given by $\mathbb{I}$, $p_\parallel$ represents the pressure along field lines and $p_\perp$ is the pressure perpendicular to them. Note that in the isotropic limit ($p_\parallel=p_\perp$), one recovers the previous MS force balance. To simplify notation we introduce a new function $\Delta\equiv(p_\parallel-p_\perp)/B^2$, which we shall treat as having no particular symmetry. Since  $p_\parallel,~p_\perp>0$, we must have $\Delta>-p_\perp/B^2$, an inequality that must be satisfied for all physically realizable solutions.

\subsection{Constructing force balance equations}
The LHS of the force balance equation was expressed in a convenient form when we considered the isotropic MS problem. We now need to find a convenient form for the divergence of the pressure tensor. \par
The force balance is rewritten in the form
\begin{equation}
    (1-\Delta)\n{j}\times\n{B}=(\n{B}\cdot\nabla\Delta) \n{B}-\frac{1}{2}J^{-2}\Delta B_\alpha^2\nabla\left(\frac{1}{B^2}\right)+\nabla p_\perp.
\end{equation}
For the LHS (see (\ref{eq:lhsFBal})),
\begin{equation*}
    \n{j}\times\n{B}=(-\Bar{\iota}A_\alpha\nabla\phi-A_\psi\nabla\psi+A_\alpha\nabla\chi)J^{-1}
\end{equation*}
where,
\begin{gather*}
    A_\alpha=(\partial_\phi+\Bar{\iota}\partial_\chi)B_\theta \\
    A_\psi=\partial_\psi B_\alpha-(\partial_\phi+\Bar{\iota}\partial_\chi)B_\psi-B_\theta\Bar{\iota}',
\end{gather*}
and $\Bar{\iota}'=\partial_\psi\Bar{\iota}$. \par
Using the covariant and contravariant forms of the magnetic field as needed, the three components of the force-balance equation may be written out straightforwardly. After minor algebraic manipulations, we obtain
\begin{gather}
    \mathrm{I}:~~~\frac{1}{B^2}(\partial_\phi+\Bar{\iota}\partial_\chi)\Delta+\frac{1}{B^4}(\partial_\phi+\Bar{\iota}\partial_\chi)p_\perp=\frac{\Bar{\iota}}{2}\Delta\partial_\chi\left(\frac{1}{B^2}\right) \\
    \mathrm{II}:~~~A_\alpha(1-\Delta)=\frac{B_\theta}{B^2}\partial_\phi p_\perp +\frac{B_\alpha-\Bar{\iota}B_\theta}{B^2}\partial_\chi p_\perp-\nonumber\\
    ~~~~~~~~~~~-\frac{B_\alpha-\Bar{\iota}B_\theta}{2}B^2\Delta \partial_\chi\left(\frac{1}{B^2}\right) \\
    \mathrm{III}:~~~A_\psi(1-\Delta)+B_\psi(\partial_\phi+\Bar{\iota}\partial_\chi)\Delta-\frac{1}{2}B_\alpha B^2\Delta\partial_\psi\left(\frac{1}{B^2}\right)+\nonumber\\
    ~~~~~~+J\partial_\psi p_\perp=0.
\end{gather}

\subsection{Expansion procedure}
\subsubsection{Equation I}
Equation I resembles its MS counterpart, yet it plays a rather different role in the construction.
\paragraph{Order $\epsilon^0$}: 
\begin{equation*}
    B_0\partial_\phi p_{\perp0}+\partial_\phi\Delta_0=0
\end{equation*}
which simply gives,
\begin{equation}
    B_0 p_{0}+\Delta_0=\mathrm{const}
\end{equation}
So the angular dependence of the anisotropy must be on axis the same as that of the perpendicular pressure. This allows $p_{\perp0}$ to be a general periodic function of $\phi$, so that neither $\Delta$ nor $p_\perp$ are constrained to be flux functions. 
\paragraph{Order $\epsilon$}:
\begin{equation*}
    B_0(\partial_\phi+\Bar{\iota}\partial_\chi)(B_0 p_1+\Delta_1)=\frac{\Bar{\iota}}{2}\Delta_0\partial_\chi B_1+B_1\Delta_0'
\end{equation*}
where the prime denotes, as usual, a derivative with respect to $\phi$. We combine the harmonic coefficients to construct SHO equations of the form,
\begin{equation*}
    (B_0p_{11}^C+\Delta_{11}^C)''+\Bar{\iota}_0^2(B_0p_{11}^C+\Delta_{11}^C)=\frac{B_{11}^C}{B_0}\left(\frac{\Bar{\iota}_0^2}{2}\Delta_0+\Delta_0''\right).
\end{equation*}
We can obtain solutions for the function $B_0p_{11}^{C/S}+\Delta_{11}^{C/S}$ from particular solutions of the equation above. Expressing $\Delta_0$ as a periodic Fourier series 
\begin{equation*}
    \Delta_0=\sum_{n=0}^\infty\left(\Bar{\Delta}_{0n}^C\cos(n\phi)+\Bar{\Delta}_{0n}^S\sin(n\phi)\right),
\end{equation*}
one may obtain explicitly, 
\begin{align}
    B_0p_{11}^C+\Delta_{11}^C&=-\eta\sum_{n=0}^\infty\frac{\Bar{\iota}_0^2-2n^2}{\Bar{\iota}_0^2-n^2}\left(\Bar{\Delta}_{0n}^S\sin n\phi+\Bar{\Delta}_{0n}^C\cos n\phi\right) \label{eq:d11C}
\end{align}
and
\begin{equation}
    B_0p_{11}^S+\Delta_{11}^S=-\Bar{\iota}_0\eta\sum_{n=0}^\infty\frac{n}{\Bar{\iota}_0^2-n^2}\left(\Bar{\Delta}_{0n}^S\cos n\phi-\Bar{\Delta}_{0n}^C\sin n\phi\right). \label{eq:d11S}
\end{equation}
Everything so far is periodic by construction, and interestingly, the sine component has a zero $\phi$ averaged value. We refer the reader to Appendix E for more details of how to construct $\Delta_{11}$.

\begin{figure}
    \centering
    \includegraphics[width=0.2\textwidth]{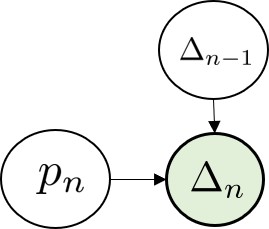}
    \caption{\textbf{Diagram for equation I$^n$.} The green shaded circle represents the function that one should solve the equation for ($\Delta_n$). The arrows indicate the leading order of the functions needed to find a solution.}
    \label{fig:aniIeqDiag}
\end{figure}

\paragraph{Higher-order $\epsilon^n$}: the harmonic structure of the equations prevails at higher order, as we show in Appendix E. Thus Equation I is used to construct $\Delta_n$ as a function of $p_n$ and lower orders at each order $n$. This constitutes a total of $n+1$ constriants (see Fig. \ref{fig:aniIeqDiag}). Note the difference with the MS construction that forces $p$ to be a flux function. \par
The equation for $\Delta_{n0}$ is a rather special one; instead of a SHO-like structure, it has the form $\partial_\phi f=g$. Though the solution is straightforward, a periodic solution for $f$ will exist if and only if we satisfy the solubility condition $\int g\mathrm{d}\phi=0$. This can, in principle, play the role of an additional constraint on the constant coefficients defining the various functions.\par

\subsubsection{Equation II}
For this equation we shall skip the $\epsilon^0$ order, as the resultant equation provides no more information than what we already have.
\paragraph{Order $\epsilon^1$}:
\begin{equation*}
    \Delta_0\partial_\chi B_1-2B_0^2\partial_\chi p_1=0.
\end{equation*}
Considering the harmonics of the expression above, it follows that
\begin{gather*}
    B_0p_{11}^S=0 \\
    B_0p_{11}^C=-\eta\Delta_0.
\end{gather*}
Thus, the leading order gives closed form expressions for the first-order pressure. The fact that the sine component of the pressure must vanish can be related to the particular choice of angular coordinates, which we have chosen to be such that $B_{11}^S=0$.

\paragraph{Order $\epsilon^2$}: 
the zeroth harmonic component reads,
\begin{align*}
    \frac{\mathrm{d}}{\mathrm{d}\phi}\left[B_{\theta20}^C(1-\Delta_0)\right]=-\frac{B_{\alpha0}}{2}\eta\Delta_{11}^S
\end{align*}
and is easily checked to satisfy the solubility condition using the form of $\Delta_{11}^S$ given in (\ref{eq:d11S}). Solving it explicitly, we obtain
\begin{align}
    &B_{\theta20}^C(1-\Delta_0)=\nonumber\\
    &=\Bar{B}_{\theta20}-B_{\alpha0}\frac{\Bar{\iota}_0\eta^2}{2}\sum_{n=0}^\infty\frac{1}{\Bar{\iota}_0^2-n^2}\left(\Bar{\Delta}_{0n}^S\sin n\phi+\Bar{\Delta}_{0n}^C\cos n\phi\right)
\end{align}
where $\Bar{B}_{\theta20}$ represents an integration constant. \par
To this order, Equation II yields two more equations corresponding to $\cos2\chi$ and $\sin2\chi$. These equations are however equivalent to part of the constraints in Equation III, so we postpone the discussion to later in the paper. 

\begin{figure}
    \centering
    \includegraphics[width=0.4\textwidth]{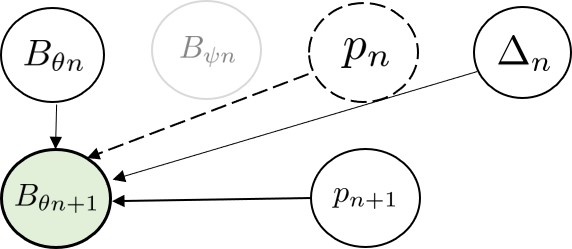}
    \caption{\textbf{Diagram for equation II$^n$.} The green shaded circle represents the function that one should solve the equation for ($B_{\theta n+1}$). The arrows indicate the leading order of the functions needed to find a solution (broken lines refer to the special case of $B_{\theta n+1,0}$).}
    \label{fig:aniIIeqDiag}
\end{figure}

\paragraph{Higher order $\epsilon^n$} 
At any order one may solve Equation II for $B_{\theta n+1}$, in a similar way as has been done for Equation I, as a function of $p_{n+1}$ and $\Delta_n$. This constitutes a total number of $n-1$ constraint equations, precisely the degrees of freedom left in $B_{\theta n}$. A diagramatic representation is given in Fig. \ref{fig:aniIIeqDiag}.\par
Note that we are discounting the equations corresponding to the two harmonics $B_{\theta nn}$. (Proof and other details on the generalisation of Equation II are presented in Appendix E.) \par
The solution structure for $B_{\theta n}$ is discussed in more detail in Appendix E. Briefly, it closely resembles that of Equation I, with a particular solution for the analogous SHO-like equations. The equation associated to $B_{\theta n0}$ also has a special character, as seen in the $\epsilon^2$ order considered earlier.  \par

\subsubsection{Equation III}
\paragraph{Order $\epsilon^0$}: 
\begin{align*}
    B_0\frac{\mathrm{d}}{\mathrm{d}\phi}\left[B_{\psi0}(1-\Delta_0)\right]=B_0B_{\alpha0}B_1p_1+B_0^2B_{\alpha0}p_2-\\
    -\frac{1}{2}B_{\alpha0}B_2\Delta_0+B_0B_{\alpha1}(1-\Delta_0)-\frac{1}{4}B_{\alpha0}B_1\Delta_1
\end{align*}
It then follows that the 0th harmonic is given by
\begin{align}
    B_0B_{\alpha0}p_{20}^C=\left[B_{\psi0}(1-\Delta_0)\right]'+B_{\alpha1}(\Delta_0-1)+\nonumber\\
    +\frac{1}{2}\frac{B_{\alpha0}}{B_0}B_{20}^C\Delta_0-\frac{1}{2}B_{\alpha0}B_{11}^Cp_{11}^C+\frac{B_{\alpha0}}{8B_0}B_{11}^C\Delta_{11}^C,
\end{align}
and the other two harmonics are given by
\begin{align}
    p_{22}^C=\frac{1}{8B_0^2}\left(4B_{22}^C\Delta_0-4B_0B_{11}^Cp_{11}^C+B_{11}^C\Delta_{11}^C\right)
\end{align}
and,
\begin{align}
    p_{22}^S=\frac{1}{8B_0^2}\left(4B_{22}^S\Delta_0+B_{11}^C\Delta_{11}^S\right).
\end{align}
The leading order of Equation III then provides us with a closed algebraic form for the pressure $p_2$, in terms of lower order pressure and anisotropy $\Delta$, as well as $B_{\psi0}$. It is important to point out that the 0th harmonic could also be viewed as an ODE for $B_{\psi0}$. However, this would involve solving an additional ODE, so it is better to adopt the given interpretation. 

\begin{figure}
    \centering
    \includegraphics[width=0.4\textwidth]{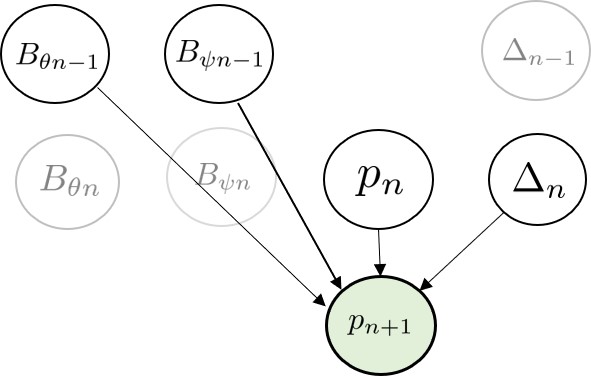}
    \caption{\textbf{Diagram for equation III$^n$.} The green shaded circle represents the function that one should solve the equation for ($p_{n+1}$). The arrows indicate the leading order of the functions needed to find a solution.}
    \label{fig:aniIIIeqDiag}
\end{figure}

\paragraph{Higher order} From the general $\epsilon^{n-2}$ case, one is able to construct $p_n$ in terms of $B_{\psi n-2}, \Delta_{n-1}$, $p_{n-1}$ and $B_{\theta n-2}$. This constitutes $n+1$ equations, fully determining $p_n$ (see Fig. \ref{fig:aniIIIeqDiag} for a diagram). We refer to Appendix E for a constructive proof.

\paragraph{}
In summary, from the anisotropic pressure force balance equation we obtain three sets of constraints, which have to be solved for different functions at different orders. These are summarised in Table \ref{tab:countingQSAnaxis}.

\begin{table}[]
    \centering \hspace*{-.5cm}
    \begin{tabular}{|c||c|c|c|c|}
        \hline
    Eqn. & Order & Solve for\dots & Nb. eqns \\ \hline\hline
    I$^n$ & $n$ & $\Delta_n$ & $n+1$ \\\hline
    II$^n$ & $n$ & $B_{\theta n}$ & $n-1$ \\
    & EVEN & $B_{\theta n0}$ & (1) \\\hline 
    III$^{n-2}$ & $n-2$ & $p_n$ & $n+1$ \\ \hline
    \end{tabular}
    \caption{\textbf{Counting of equations and degrees of freedom.} Each column shows: the label of equations, the order of expansion, what the equations are solved for, and the number of consraints (or independent equations) they amount to. We have included the slightly different nature of the $B_{\theta n0}$ function.}
    \label{tab:countingQSAnaxis}
\end{table}

\subsection{Complete near-axis construction for anisotropic pressure}
To complete successfully the order-by-order construction of a solution, a detailed analysis of the conjuction of the magnetic and anisotropic force balance equations is needed. To begin, we undertake a tentative count of equations and degrees of freedom to see if the system is overdetermined as in the case of isotropic pressure. \par
From previous sections, we are able to find equations for each of the functions $X,~Y,~Z,~B_\psi$ (magnetic equations),$~B_\theta,~p$ and $\Delta$ (force balance equations) at every order, leaving only $Y_{n0}$ and $B_{\psi n0}$ functions free. To remind the reader, the relevant equations are labelled $J$, $C_\perp$, $C_b$, I, II and III. This suggests that in the anisotropic pressure case there is no overdetermination, and the near-axis construction is not limited to the first few orders. In this case, the expansion can in principle be carried to arbitrarily high order, as it seems that there are not only enough equations to solve for all of the functions, but even some unconstrained free functions. \par
This tentative counting process, however, should be taken with a grain of salt because no explicit prescription of how to proceed from one order, $n$, to another, $n+1$, has been given. This requires both an understanding of how different functions get involved at different orders, furthermore, how they are mutually related. This was straightforward in the context of the MS equilibrium (see Fig. \ref{fig:ConstructNAMS}), but the complexity of the present problem requires additional attention. \par
\begin{figure*}
    \includegraphics[width=0.7\textwidth]{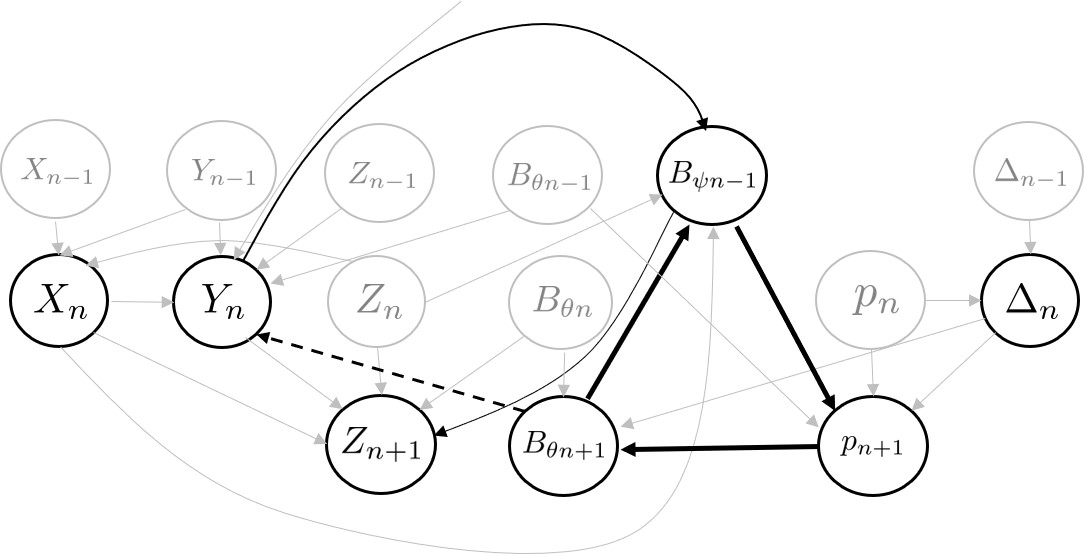}
    \caption{\textbf{Function dependency from equations for anisotropic expansion.} The diagram shows the function dependency as governed by the equations obtained in Tabs. \ref{tab:countingQSJCaxis} and \ref{tab:countingQSAnaxis}. The darker circles represent the order of the functions that come to play at the same time according to those equations. The arrows flow from the functions needed to be known to those that are being solved for. The darker lines show the main difficulty of this construction, what we refer to as the \textit{loop}. The broken line represents a relation that only holds every other order through the $B_{\theta n0}$ component. }
    \label{fig:loopFunAna}
\end{figure*}

As a first step, we represent all the functional relations with the help of a diagram, shown in Fig. \ref{fig:loopFunAna}.
There is one especially noteworthy feature in the diagram: a \textit{loop} highlighted with thicker black arrows. This closed cycle indicates that the three functions $p_{n+1},~B_{\theta n+1}$ and $B_{\psi n-1}$ depend on each other through the set of equations $C_\perp$, II and III. As it stands, the construction is not explicit: the loop needs to be undone. The standard way to \textit{un-loop} these equations would be to eliminate two of the functions from one of the equations in terms of the other. Such an equation will be referred to as the \textit{looped} equation (referred to as $\tilde{\mathrm{II}}$, given that this equation results from the substitution into II). With $\tilde{\mathrm{II}}$ in terms of only one of the functions, it may be solved, and eventually used to construct the other two functions.\par
However, and as shown systematically in Appendix F, the \textit{looped} equations end up containing no terms with the higher-order functions $p_{n+1},~B_{\theta n+1}$ and $B_{\psi n-1}$. This precise cancellation occurs to all orders, casting these equations in the role of additional constraints on the remaining functions to order $n$ and other order $n-1$. (The $m=0$ case is an exception in that it does contain $B_{\theta n 0 }$ even after the looping.)  \par
\begin{figure}
    \includegraphics[width=0.4\textwidth]{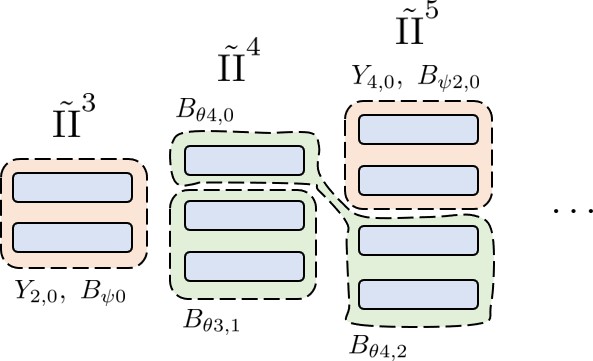}
    \caption{\textbf{Solving the \textit{looped} equations $\tilde{\mathrm{II}}$.} Each of the rounded rectangles in the diagram represents a constraint equation, ordered in columns corresponding to the \textit{looped} forms of Equation II, with the lowest harmonics on top of each column. The function labels represent the functions to be solved for. There are two color schemes: the reddish that we refer to as self-consistent equation ($\tilde{\mathrm{II}}_\mathrm{SC}$), and the greenish that are solved for the $B_\theta$ functions.}
    \label{fig:loopEqDiag}
\end{figure}

The question now is how these \textit{looped} equations are to be accommodated. Given that at every order $B_{\theta n}$ drops out, we should employ the looped equations to solve for $B_{\theta n-1}$. This does not exhaust all the available constraints though. Let us look then into the first couple of orders to see how these extra constraints present themselves. To help the discussion, we present in Fig. \ref{fig:loopEqDiag} a schematic description of how these loop equations are used. \par
The first time we need to deal with a loop is $n=2$. In that case, the free functions available to the \textit{looped} equations are $Y_{20},~B_{\psi 0}$ and $\Delta_0$ (or $p_0$, depending on interpretation), in addition to the curvature $\kappa$ and torsion $\tau$ describing the magnetic axis. The two looped equations $\tilde{\mathrm{II}}^3$ may be solved for $Y_{20}$ and $B_{\psi 0}$, in a self-consistent manner that reminds us of its MS counterpart.\cite{garrenboozer1991b} Doing so postpones the solution to $B_{\theta 3},~p_3$ and $B_{\psi1}$ to the next order, thus changing the order in which functions appear in the construction. Note the important difference between the standard procedure\cite{garrenboozer1991b} and this one: here the axis is left unconstrained. This additional freedom will be the focus of some numerical work in Part II of our sequence. \par
In the next order, we have three self-consistent \textit{looped} constraints from $\tilde{\mathrm{II}}^4$. The larger harmonics do not include any reference to $B_{\theta 4}$ after the equations are \textit{looped}, and thus ought to be solved for $B_{\theta3,1}$. To that end, the equation has to be rewritten with the help of III and $C_\perp$, all in terms of $B_\theta$. Once the solutions for $B_{\theta 31}^{C/S}$ are found, one may then express $p_3$ and $B_{\psi1}$ in a closed form using the appropriate form of III$^1$ and $C_\perp^2$ respectively. As a final step, the non-harmonic component of $\tilde{\mathrm{II}}^4$ (which we call $\tilde{\mathrm{II}}_0^4$) is solved explicitly for $B_{\theta 40}$, necessary to find $Y_3$. This latter step could be mixed with the previous steps, complicating the solution construction, but not altering the procedure in any fundamental way. No exploration of these resulting equations nor a study of the existence of solutions is attempted here. Our tentative construction is thus not a rigorous proof of the existence of global QS solutions, but rather should be viewed as a systematic procedure for construction of solutions by power-series exapansion.  \par
\begin{figure}
    \includegraphics[width=0.5\textwidth]{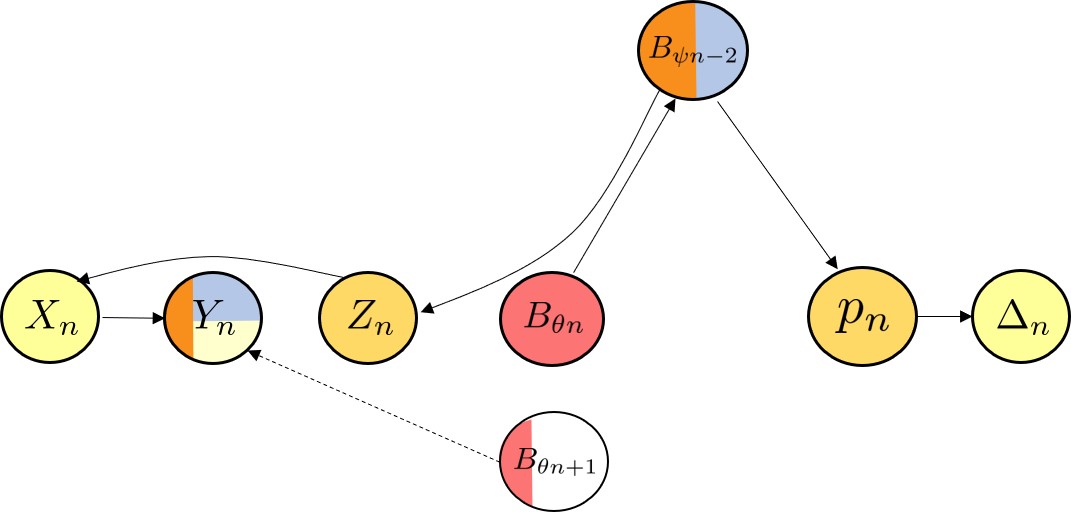}
    \caption{\textbf{Consistent order of functions.} Order of functions that ought to be simultaneously solved for. The color code represents the order in which the functions are solved for: red - first to be solved using the \textit{looped} form of II, dark orange, bright orange, yellow and pale yellow. The blue color represents the introduction of free functions at even orders that are to be solved self consistently. The dashed arrow and reddish colored part of $B_{\theta n+1}$ represent that for odd orders, the 0 harmonic term of the next order is to be solved simultaneously as well. The arrows show the mutual dependencies, originating from $B_{\theta n}$.}
    \label{fig:ConstructNAAnis}
\end{figure}

Let us now extend the procedure above to higher orders:
\begin{itemize}
    \item if the order $n$ is odd, then the $\tilde{\mathrm{II}}^{n+1}$ equations should be used (in conjuction with III and $C_\perp$) to solve for $B_{\theta n}$ and $B_{\theta n+1 ,0}$. 
    \item if the order $n$ is even, then two of the $\tilde{\mathrm{II}}^{n+1}$ equations should be used to self-consistently solve for $Y_{n0}$ and $B_{\psi n-2 ,0}$ (similarly to the procedure followed for $n=2$, $\tilde{\mathrm{II}}_\mathrm{SC}^{n+1}$), and the other remaining $n-2$ equations should be used for solving $B_{\theta n}$ (excluding $B_{\theta n,0}$ which had been solved in the previous order).
\end{itemize} 
Once solved, one may then explicitly construct $B_{\psi n-2}$, to continue with  $Z_n$ and $p_n$, following $\Delta_n$ and $X_n$. With this knowledge and $B_{\theta n+1,0}$, finally $Y_n$ can be completely solved for. This flow of the construction, and the order at which functions appear is presented in Fig. \ref{fig:ConstructNAAnis}. Table \ref{tab:eqConstructNAAnis} presents the equations necessary to find solution to the functions at each order. As usual, each cell indicates the equations that need to be used to solve for the order and function identified with the position in the Table.
\begin{table}[]
    \centering
    \includegraphics[width=0.5\textwidth]{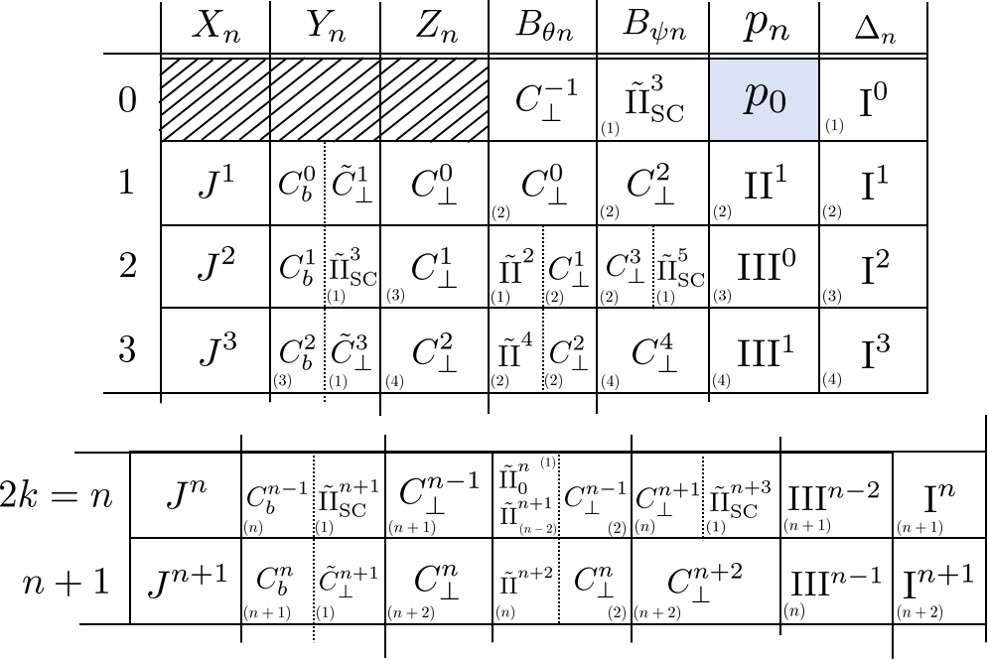}
    \caption{\textbf{Anisotropic near-axis construction,} Each cell represents the equations that need to be solved to find a function (column) at a given order (row). The equation labels are as follow: $J$ for the Jacobian equation, $C_\perp$ and $C_b$ for the co(ntra)variant perpendicular and $b_0$ equations, and I, II and III for the force balance equations. According to the discussion on the \textit{loop}, the notation$\Tilde{\mathrm{II}}$ refers to the \textit{looped} version of the equation, with the underscript SC referring to the self-consistent equation to be solved for $Y_{n0}$ and $B_{\psi n-2,0}$. The numbers in parenthesis denote the number of constraint equations (no number is given in some cells in which the number of unknowns equals that of constraints). The blue color shows the only free function left in the construction apart from the curvature and torsion of the axis.}
    \label{tab:eqConstructNAAnis}
\end{table}
\par
Using Table \ref{tab:eqConstructNAAnis}, the counting of equations and degrees of freedom is straightforward. It follows that with Table \ref{tab:eqConstructNAAnis} perfectly tiled by equations, the construction is neither overconstrained nor underconstrained. In fact, there are precisely the same number of constraint equations that need to be solved at each order as the number of functions that appear anew. It is easy to check that the equation number in the table add up to the appropriate number, as indicated by an equation number reference table (Table \ref{tab:countingEqNum}). \par
\begin{table}[]
    \centering \hspace*{-.5cm}
    \begin{tabular}{|c|c||c|c|}
        \hline
        Eqn. & Nb. eqns&Eqn. & Nb. eqns \\ \hline\hline
    I$^n$ & $n+1$ & $J^n$ & $n+1$\\
    II$^n$ & $n-1$ & $C_b^n$ & $n+1$\\
    III$^{n-2}$ & $n+1$ & $C_\perp^n$ & $2(n+2)$\\\hline
    \end{tabular}
    \caption{\textbf{Number of constraint equations.} Number of total independent constraints associated to each one of the relevant equations.}
    \label{tab:countingEqNum}
\end{table}
In summary, as far as the counting goes, by relaxing the assumption of MS equilibrium with isotropic pressure, the near-axis expansion avoids the problem of overdetermination. Assuming equilibrium force-balance of QS fields with anisotropic pressure, we have shown that it is in principle possible to continue the near-axis expansion to orders higher than two. While this may suggest that it is possible, \textit{in principle} to continue the process to arbitrarily higher order, physicality aside, we have not explicitly proved that the equations to be solved at higher order, especially the looped ones, have a solution. \par
Despite not being able to use the present construction as a definitive proof of the existence of a global solution to quasisymmetry, our work suggests that there is likely a way to construct globally QS fields. This is qualitatively opposite to accepted wisdom, prompted by [\onlinecite{garrenboozer1991b}], is that whereas QS solutions can be realized on a particular flux surface, they do not exist globally (except in cases of continuous symmetry such as axisymmetry). \par
Following our construction procedure, solutions may in principle be built in a consistent way to arbitrarily high order for some given free form of $p_0$. In fact, there seems to be no fundamental limitation as to how arbitrarily close the pressure profile may be to isotropy near axis.

\section{Conclusion}
In this paper, a near-axis expansion procedure for the construction of a QS magnetic field in general force balance has been presented. The developments presented here were made possible by separating the mathematical conditions of quasi-symmetry from conditions of force balance, enabling us to treat the two cases of scalar and anisotropic pressure separately.  \par
For the scalar pressure case, we show that the procedure reduces to the well known form in [\onlinecite{garrenboozer1991b}]. In particular, we obtain the well-known result that the construction leads to the problem of overdetermination. However, the introduction of anisotropic pressure appears to alleviate the problem of overdetermination, leading to the possibility that QS solutions may be constructed to arbitrarily high order. It also appears that isotropy could in principle be approximated arbitrarily closely without running into the problem overdetermination, idea explored further in the companion Part II paper. This suggests that, unlike popular belief, it might be possible to find globally quasisymmetric field solutions. \par
In an ensuing paper (Part II), we will apply this construction procedure to numerically find QS field solutions that have been impossible so far. In particular, we shall explore in depth the problem of QS stellarators with a circular axis as a proof of principle.

\section*{Appendix A. Physical confinement properties of weakly quasisymmetric fields}
We discuss some of the basic confinement properties that result from \textit{weak} quasisymmetry\cite{rodriguez2020}\footnote{The nomenclature \textit{weak quasisymmetry}  was suggested by D. Ginsberg (private communication, 2020) for the definition of quasisymmetry in [\onlinecite{rodriguez2020}], to distinguish it from \textit{strong quasisymmetry}, in the more constraining definition of [\onlinecite{burby2019}].}. By \textit{weak} quasisymmetry we mean a field that satisfies,
\begin{gather}
    \n{u}\cdot\nabla B=0 \\
    \n{B}\times\n{u}=\nabla \psi  \\
    \nabla\cdot\n{u}=0,
\end{gather}
where $\mathbf{u}$ represents a non-zero vector field that points in the direction of constant $B$. For a detailed discussion we refer the reader to [\onlinecite{rodriguez2020}]. \par
The first important property that follows from \textit{weak} quasisymmetry is that the single particle motion has an approximatedly conserved quantity $\Bar{p}$,
\begin{equation}
    \Bar{p}=-\frac{1}{\epsilon}\psi+v_\parallel\n{u}\cdot\n{b},
\end{equation}
where $v_\parallel$ is the parallel velocity of the charged particle. This conservation holds to order $O(\epsilon)$, where $\epsilon=\rho/L$, $\rho$ is the particle gyroradius and $L$ is some characteristic macroscopic length of the system. Then, particles are restricted to remain close to flux surfaces of constant $\psi$ to leading order. \par
This approximate conservation of canonical momentum, though important for neoclassical transport, is rather weak on its own, given that it allows for collisionless departures of particles in times on the order of $1/\Omega\epsilon$, where $\Omega$ is the gyrofrequency. It is, however, straightforward to show that, the \textit{weak} requirement of QS also implies the alignment of surfaces of constant approximate second adiabatic invariant with flux surfaces, suggesting collisionless confinement at longer times. To prove this, and dropping the electric potential, we write
\begin{align*}
    J_2=&\oint v_\parallel\mathrm{d}l=\oint\sqrt{2(E-\mu B)}\mathrm{d}l\\
    =&\oint\sqrt{2(E-\mu B)}\frac{\mathrm{d}B}{\n{B}\cdot\nabla B},
\end{align*}
where $E$ is the energy of the particle and $\mu$ its associated magnetic moment. Note that we have changed the integration variable to the magnetic field magnitude along field lines. From weak quasisymmetry\cite{rodriguez2020}, it follows that $\n{B}\cdot\nabla B=f(\psi,B)$, which implies that the $J_2$ integral can be written explicitly in the form,
\begin{equation*}
    J_2=\oint g(\psi,B,E,\mu)\mathrm{d}B=J_2(\psi,E,\mu),
\end{equation*}
because the endpoints depend only on $\mu$ and $E$. Thus, we have shown that surfaces of constant $J_2$ are aligned with flux surfaces. 

\section*{Appendix B. Existence of generalized Boozer representation for QS fields}
We show that it is always possible to find a Boozer-like straight-field-line coordinate system in the sense presented in Section I of this paper. \par
Let us start by constructing a coordinate system with a Jacobian that depends on space only through $\psi$ and $B$ from given straight-field-line coordinate system $\{\psi,\theta,\phi\}$. We can write $\n{B}=\nabla\psi\times\nabla\theta+\iota\nabla\phi\times\nabla\psi$, which is always possible as long as magnetic flux surfaces exist, so $\psi$ is a single- valued function and $\nabla\psi\neq0$. \par
There is a whole family of straight field line transformations that leaves $\n{B}$ unchanged. The map generated by $\omega$,
\begin{gather*}
\theta=\theta'+\iota\omega \\
\phi=\phi'+\omega,
\end{gather*}
where $\omega=\omega(\psi,\theta,\phi)$ is some well behaved periodic function (which thus preserves the poloidal/toroidal nature of the angles). Under such a transformation the Jacobian transforms according to the relation,
\begin{align}
    J'^{-1}=J^{-1}-\n{B}\cdot\nabla\omega. \label{jacTrans}
\end{align}
Now we require the newly defined coordinate system $\{\psi,\theta',\phi'\}$ to have a Jacobian $J'$ which depends only on space through $B$ and $\psi$. The question is whether there always exists some $\omega$ so that this is true. The equation that we need to solve is,
\begin{equation*}
    \n{B}\cdot\nabla\omega=\frac{1}{J}-\frac{1}{J'}.
\end{equation*}
For this equation to have a single-valued solution for $\omega$, 
\begin{equation*}
    \langle\n{B}\cdot\nabla\omega\rangle=0=\left\langle\frac{1}{J}\right\rangle-\left\langle\frac{1}{J'}\right\rangle,
\end{equation*}
where $\langle\dots\rangle$ represents the flux surface average. Explicitly, the flux surface averaging may be written (see [\onlinecite{Helander2014}]),
\begin{equation*}
    \left\langle\frac{1}{J}\right\rangle=\frac{1}{V'}\int_0^{2\pi}\mathrm{d}\alpha\int\frac{\mathrm{d}l}{B}\frac{1}{J}
\end{equation*}
where $V'=\langle1\rangle V'$, and $\alpha$ labels field lines. The second integral is to be taken along the magnetic field lines once around the toroidal direction, and may be parametrised conveniently by $B$. As there could in principle be multiple points along the field line with the same value $B$, we should formally keep an additional label to distinguish them, but we omit this for simplicity. \par
This change of variables yields, assuming isolated $\mathbf{B}\cdot\nabla B=0$ extrema,
\begin{equation*}
    \left\langle\frac{1}{J}\right\rangle=\frac{1}{V'}\int_0^{2\pi}\mathrm{d}\alpha\int\frac{\mathrm{d}B}{\n{B}\cdot\nabla B}\frac{1}{J}.
\end{equation*}
We would now like to change the order of the integrals. Because lines of constant $B$ do not intersect and are nowhere tangent to magnetic field lines so as to avoid singular points on flux surfaces that pass through points where $\nabla\psi=0$, the integrals may be made to commute by changing the integration limits carefuly so as to tile the whole flux surface. Furthermore, because we are concerned with QS fields, the expression $\n{B}\cdot\nabla B=f(\psi,B)$. This implies that
\begin{equation*}
    \left\langle\frac{1}{J}\right\rangle=\frac{1}{V'}\int\frac{\mathrm{d}B}{\n{B}\cdot\nabla B}\int_{\alpha_0(B)}^{\alpha_f(B)}\mathrm{d}\alpha \frac{1}{J}.
\end{equation*} 
Similarly, we can write,
\begin{equation*}
    \left\langle\frac{1}{J'}\right\rangle=\frac{1}{V'}\int\frac{\mathrm{d}B}{\n{B}\cdot\nabla B}\frac{\alpha_f(B)-\alpha_0(B)}{J'}.
\end{equation*}
Choose a Jacobian $J'$ of the form,
\begin{equation}
        \frac{1}{J'}=\frac{1}{\alpha_f(B)-\alpha_0(B)}\int_{\alpha_0(B)}^{\alpha_f(B)}\mathrm{d}\alpha\frac{1}{J},
\end{equation}
keeping both $B$ and $\psi$ constant, so that by construction it only depends on $B$ and $\psi$. Then a single-valued solution $\omega$ can be found by solving the magnetic equation
\begin{equation}
    \n{B}\cdot\nabla\omega=J^{-1}\left[1-\frac{J}{\alpha_f(B)-\alpha_0(B)}\int_{\alpha_0(B)}^{\alpha_f(B)}\mathrm{d}\alpha J^{-1}\right].
\end{equation}
It thus follows, by construction, that beginning from a well-behaved straight-field-line coordinate system for a QS field, we can always find a coordinate system in which the Jacobian depends on $B$ and $\psi$ only.\par
For this transformation to be well-behaved, we must impose the requirement that the Jacobian is invertible at any point in the region of interest. It is sufficient to require $\n{B}\cdot\nabla\omega\neq J^{-1}$. Given our construction, and assuming smoothness, this implies the requirement that,
\begin{equation*}
    J\left[\frac{1}{\alpha_f(B)-\alpha_0(B)}\int_{\alpha_0(B)}^{\alpha_f(B)}\mathrm{d}\alpha J^{-1}\right]\neq0.
\end{equation*} 
Provided the original straight-field-line coordinate system is well-behaved, with $J\neq0$ over flux surfaces, then smoothness implies that $J>0$ or $J<0$ everywhere over the surface. The averaging over field lines should preserve the sign of $J$ (since $J$ itself does not change sign). Thus the construction given avoids the coordinate transformation singularity.\par
In order to complete the construction of the generalizd Boozer coordinates, there remains only one additional step. We need to show that the Jacobian of such a coordinate system may be written in the form $J=B_\alpha(\psi)/B^2$. Consider an additional coordinate transformation induced by $\tilde{\omega}$, that is,
\begin{equation*}
    \n{B}\cdot\nabla\tilde{\omega}=\frac{B^2}{B_\alpha}-\frac{B^2}{B_{\alpha'}},
\end{equation*} 
where $J=B_\alpha/B^2$ from (\ref{eqn:jacCoord}).
To see whether a $\tilde{\omega}$ can be found such that $B_\alpha$ is a flux function, we need to analyze the solubility condition of the magnetic equation. Provided $B_\alpha=\langle B^2\rangle/\langle B^2/B_{\alpha'}\rangle$, such a coordinate transformation does exist, and in fact preserves the good behaviour of the Jacobian.

\section*{Appendix C. Second order Jacobian equation}
To order $\epsilon^2$, the Jacobian equation reads
\begin{align*}
   J^2:~~~~~ 2\left(\frac{\mathrm{d}l}{\mathrm{d}\phi}\right)^2\kappa X_2=-2B_{\alpha0}B_{\alpha1}B_0-B_{\alpha0}^2B_{2}+\\
   +\left(\frac{\mathrm{d}l}{\mathrm{d}\phi}\right)^2X_1^2\kappa^2+2\left(\frac{\mathrm{d}l}{\mathrm{d}\phi}\right)Z_2'
    +\\
    +\left(\frac{\mathrm{d}l}{\mathrm{d}\phi}Y_1\tau+X_1'+\Bar{\iota}_0\Dot{X}_1\right)^2+\\
    +\left(-\frac{\mathrm{d}l}{\mathrm{d}\phi}X_1\tau+Y_1'
    +\Bar{\iota}_0\Dot{Y}_1\right)^2+2\left(\frac{\mathrm{d}l}{\mathrm{d}\phi}\right)\Bar{\iota}_0\Dot{Z}_2
\end{align*}
where the prime denotes a derivative with respect to $\phi$ and a dot a derivative with respect to $\chi$. For completeness, we provide below the explicit form for the components of $X_2$: 
\begin{align*}
X^C_{2,0}{}&= \frac{1}{4 l^2 \kappa {}}\left[-2 B_{\alpha0}^2 B^C_{2,0}+l^2 \kappa {}^2 X^C_{1,1}{}{}^2+l^2 \tau {}^2 X^C_{1,1}{}{}^2+\right.\\
&\left.+l^2 \tau {}^2 Y^C_{1,1}{}{}^2+l^2 \tau {}^2 Y^S_{1,1}{}{}^2+2 l \tau {} Y^C_{1,1}{} X^C_{1,1}{}'-2 l \tau {} X^C_{1,1}{} Y^C_{1,1}{}'-\right.\\
&\left.-2 \Bar{\iota} _0 \left(2 l \tau {} X^C_{1,1}{} Y^S_{1,1}{}-Y^S_{1,1}{} Y^C_{1,1}{}'+Y^C_{1,1}{} Y^S_{1,1}{}'\right)+4 l Z^C_{2,0}{}'+\right.\\
&\left.+\Bar{\iota} _0^2 \left(X^C_{1,1}{}{}^2+Y^C_{1,1}{}{}^2+Y^S_{1,1}{}{}^2\right)+X^C_{1,1}{}'{}^2+\right.\\
&\left.+Y^C_{1,1}{}'{}^2+Y^S_{1,1}{}'{}^2-4 B_0 B_{\alpha1} B_{\alpha0}\right]
\end{align*}
\begin{align*}
 X^C_{2,2}{}&= \frac{1}{4 l^2 \kappa {}}\left[-2 B^C_{2,2} B_{\alpha0}^2+l^2 \kappa {}^2 X^C_{1,1}{}{}^2+l^2 \tau {}^2 X^C_{1,1}{}{}^2+\right.\\
 &\left.+l^2 \tau {}^2 Y^C_{1,1}{}{}^2-l^2 \tau {}^2 Y^S_{1,1}{}{}^2+X^C_{1,1}{}'{}^2+Y^C_{1,1}{}'{}^2-Y^S_{1,1}{}'{}^2-\right.\\
 &\left.\Bar{\iota} _0^2 \left(X^C_{1,1}{}{}^2+Y^C_{1,1}{}{}^2-Y^S_{1,1}{}{}^2\right)+2 l \tau {} Y^C_{1,1}{} X^C_{1,1}{}'-\right.\\
 &\left.-2 l \tau {} X^C_{1,1}{} Y^C_{1,1}{}'+2 \Bar{\iota} _0 \left(4 l Z^S_{2,2}{}+Y^S_{1,1}{} Y^C_{1,1}{}'+Y^C_{1,1}{} Y^S_{1,1}{}'\right)+\right.\\
 &\left.+4 l Z^C_{2,2}{}'\right]
\end{align*}
\begin{align*}
X^S_{2,2}{}&= \frac{1}{2 l^2 \kappa {}}\left[-B_{\alpha0}^2 B^S_{2,2}+l^2 \tau {}^2 Y^C_{1,1}{} Y^S_{1,1}{}-\Bar{\iota} _0 \left(4 l Z^C_{2,2}{}+\right.\right.\\
&\left.\left.+X^C_{1,1}{} X^C_{1,1}{}'+Y^C_{1,1}{} Y^C_{1,1}{}'-Y^S_{1,1}{} Y^S_{1,1}{}'\right)+l \tau {} Y^S_{1,1}{} X^C_{1,1}{}'-\right.\\
&\left.-l \tau {} X^C_{1,1}{} Y^S_{1,1}{}'+2 l Z^S_{2,2}{}'-\iota _0^2 Y^C_{1,1}{} Y^S_{1,1}{}+Y^C_{1,1}{}' Y^S_{1,1}{}'\right]
\end{align*}
Here, for economy of notation, we have used the shorthand $l$ for $\mathrm{d}l/\mathrm{d}\phi$. One could rewrite these expressions by factoring out terms in the RHS, but we shall not be concerned with this here.

\section*{Appendix D. Co(ntra)variant generalizations}
Let us start by constructing the solution for $Z_{n+1}$ using the order $\epsilon^n$ equations $C_\perp^n$. The $Z$ functions of interest are an order higher than that of the equation, meaning that only terms that include flux $\psi$ partial derivatives and $Z$ (with no other zeroth order vanishing function) will possibly include $Z_{n+1}$. Looking at the $Z_{n+1}$ terms then, we obtain
\begin{gather*}
    C_\kappa^n:~~~~-B_{\alpha0}\left[\Bar{Y}_1\frac{n+1}{2}Z_{n+1,m}-m\Bar{Z}_{n+1,m}\frac{1}{2}Y_1\right]=\dots \\
    C_\tau^n:~~~~-B_{\alpha0}\left[-\Bar{X}_1\frac{n+1}{2}Z_{n+1,m}+m\Bar{Z}_{n+1,m}\frac{1}{2}X_1\right]=\dots
\end{gather*} 
where the dots represent terms other than $Z_{n+1}$ and the barred functions suggest interchange of the sine and cosine coefficients with a change of sign in the cosine term (as a result of a $\chi$ derivative). All the terms in the dots are necessarily of lower order than $n+1$ given that $B_{\theta0}$, $B_{\theta1}$ and $Z_1$ are all zero. Given the form of the equations above, the correct construction for $Z_{n+1}$ at any order consists of adding $X_1 C_\kappa^n+Y_1 C_\tau^n$, so that
\begin{equation*}
    B_{\alpha0}\sqrt{B_0}(n+1)Z_{n+1,m}=\dots
\end{equation*}
This constitutes an explicit, generalised construction of $Z$.\par
Now, let us move on to show that the largest harmonics of $B_{\theta n}$ do in fact vanish to any order $n$. Take $C_\kappa^n$ as an example. Terms with partial $\psi$ derivatives are the only candidates to include the $B_{\theta n+1}$ terms,
\begin{align*}
    -B_{\alpha0}\sum_{m=1}^{n-1}(\partial_\chi Y_{n+1-m}\partial_\psi Z_m-\partial_\chi Z_m\partial_\psi Y_{n+1-m})-\\
    -B_{\theta n+1}\partial_\psi Y_1\frac{\mathrm{d}l}{\mathrm{d}\phi}.
\end{align*} 
Because we are interested in the largest harmonic possible, this requires \textit{saturation} of each of the functions, that is, if a function appears at an order $m$, then we must consider the $m$-th harmonic coefficient for it (this definition of \textit{saturation} will be employed again later). Otherwise, it would not contribute to the largest harmonic. Let us see what the implications of this are in the `commutation term' that appears in the summation of the expression above. For a given $m$,
\begin{align*}
    \partial_\chi& Y_{n+1-m}\partial_\psi Z_m-\partial_\chi Z_{n+1-m}\partial_\psi Y_m=\\
    &=\frac{m(n+1-m)}{2}\left(\bar{Y}_{n+1-m}Z_m-\bar{Z}_mY_{n+1-m}\right)=\\
    &\propto\left[\cos(n+1)\chi\left(Y_{n+1-m}^SZ_m^C+Y_{n+1-m}^CZ_m^S-\right.\right.\\
    &\left.\left.-Y_{n+1-m}^CZ_m^S-Y_{n+1-m}^SZ_m^C\right)+\sin(n+1)\chi\dots\right]
\end{align*}
where we used $\cos a\cos b=[\cos(a+b)+\cos(a-b)]/2$ and similar multiple angle formulas to obtain the relevant $n+2$ harmonics. The commutation terms vanish exactly, and therefore the only contribution to the $n+2$ harmonic comes from the $B_{\theta n+1}$ term. In particular, from $C_\kappa$ and $C_\tau$,
\begin{gather*}
    -B_{\theta n+1,n+1}\partial_\psi Y_1\frac{\mathrm{d}l}{\mathrm{d}\phi}=0 \\
    B_{\theta n+1,n+1}\partial_\psi X_1\frac{\mathrm{d}l}{\mathrm{d}\phi}=0
\end{gather*}
And thus, 
\begin{equation}
    B_{\theta n+1,n+1}=0.
\end{equation}
The trivial solution takes the place of four constraint equations. \par
To accommodate the remaining constraint equations in $C_\perp$, it is convenient to introduce an alternative form to equations (\ref{eq:Ck}) and (\ref{eq:Ct}). 
Taking the dot product of the original co(ntra)variant equation (\ref{eq:co(ntra)variant}) with $\partial\n{x}/\partial\psi$ and $\partial\n{x}/\partial\chi$ respectively, we obtain
\begin{align}
    B_\psi J=&\partial_\psi X\left(\partial_\phi X+\tau Y\frac{\mathrm{d}l}{\mathrm{d}\phi}+Z\kappa\frac{\mathrm{d}l}{\mathrm{d}\phi}+\Bar{\iota}\partial_\chi X\right)+\nonumber\\
    &+\partial_\psi Y\left(\partial_\phi Y-X\tau\frac{\mathrm{d}l}{\mathrm{d}\phi}+\Bar{\iota}\partial_\chi Y\right)+\nonumber\\
    &+\partial_\psi Z\left(\partial_\phi Z-X\kappa\frac{\mathrm{d}l}{\mathrm{d}\phi}+\frac{\mathrm{d}l}{\mathrm{d}\phi}+\Bar{\iota}\partial_\chi Z\right) \label{eq:alterC1}
\end{align}
and
\begin{align}
    B_\theta J=&\partial_\chi X\left(\partial_\phi X+\tau Y\frac{\mathrm{d}l}{\mathrm{d}\phi}+Z\kappa\frac{\mathrm{d}l}{\mathrm{d}\phi}+\Bar{\iota}\partial_\chi X\right)+\nonumber\\
    &+\partial_\chi Y\left(\partial_\phi Y-X\tau\frac{\mathrm{d}l}{\mathrm{d}\phi}+\Bar{\iota}\partial_\chi Y\right)+\nonumber\\
   & +\partial_\chi Z\left(\partial_\phi Z-X\kappa\frac{\mathrm{d}l}{\mathrm{d}\phi}+\frac{\mathrm{d}l}{\mathrm{d}\phi}+\Bar{\iota}\partial_\chi Z\right). \label{eq:alterC2}
\end{align}
Let us look at leading order forms of the equations (order $\epsilon^{n-1}$ for the $B_\psi$ and $\epsilon^{n+1}$ for the $B_\theta$), which we may write,
\begin{gather*}
    B_{\theta n+1,m}B_{\alpha0}B_0=m\frac{\mathrm{d}l}{\mathrm{d}\phi}\Bar{Z}_{n+1,m}+\dots \\
    B_{\psi n-1,m}B_{\alpha0}B_0=\frac{n+1}{2}\frac{\mathrm{d}l}{\mathrm{d}\phi}Z_{n+1,m}+\dots
\end{gather*}
where the dots represent lower order terms. It is a straightforward operation to eliminate the $Z_{n+1,m}$ terms, and to write,
\begin{gather}
    B_{\theta n+1,m}^C=\frac{2m}{n+1}B_{\psi n-1,m}^S+\dots \label{eq:CpBpSBtC}\\
    B_{\theta n+1,m}^S=-\frac{2m}{n+1}B_{\psi n-1,m}^C+\dots \label{eq:CpBpCBtS}
\end{gather}
with $m=n-1,n-3\dots\in\mathbb{N}$. Thus, we have constructed $n$ constraint equations which relate functions $B_\psi$ to $B_\theta$ and other lower order functions. A recount of the relevant functions can be done by looking at the original equation. This shows that we may express $B_{\psi n-1}$ in terms of $B_{\theta n+1},~B_{\psi n-2},~X_{n},~Y_{n}$ and $Z_{n}$, as indicated in the main text. \par
An objection could be raised that we have not made use solely of the $C_\perp$ equations in arriving at this $B_\psi$ construction, as we have in fact used the alternative form of the equations (\ref{eq:alterC1}) and (\ref{eq:alterC2}). This is technically true: the equations constitute different projections, and as such, constitute alternative linear combinations of the equations. Alternatively, we could have considered taking the form of $Z_{n+1}$ constructed explicitly and substituted it back into one of the equations of the $C_\perp$ set. This procedure is, however, significantly more convoluted and does not change the final outcome. 
\par
The only remaining generalization is the construction of $Y_{n+1}$ from the $n$-th order form of $C_b^n$. Focusing on the $Y_{n+1}$ coefficient, we may write
\begin{equation}
    -\frac{B_{\alpha0}}{2}\left[\Bar{X}_1(n+1)Y_{n+1}-mX_1\Bar{Y}_{n+1,m}\right]=\dots,
\end{equation}
where the dots represent some combination of functions $B_{\psi n-2},~X_{n+1},~B_{\theta n}$ and $Z_n$. The equation becomes simply an algebraic system of equations for the harmonic coefficients of $Y_{n+1}$. 

\section*{Appendix E. Anisotropic force balance equations: derivations}
\subsection*{Equation I. Harmonic structure}
Let us start with the order $\epsilon^1$. The original harmonic coefficients from the equation take the form,
\begin{gather*}
    (B_0p_{11}^C+\Delta_{11}^C)'+\Bar{\iota}_0(B_0p_{11}^S+\Delta_{11}^S)=\frac{B_{11}^C}{B_0}\Delta_0' \\
    (B_0p_{11}^S+\Delta_{11}^S)'-\Bar{\iota}_0(B_0p_{11}^C+\Delta_{11}^C)=-\frac{\Bar{\iota}_0}{2}\frac{B_{11}^C}{B_0}\Delta_0.
\end{gather*} 
Substituting one into the other, one obtains,
\begin{equation*}
    (B_0p_{11}^C+\Delta_{11}^C)''+\Bar{\iota}_0^2(B_0p_{11}^C+\Delta_{11}^C)=\frac{B_{11}^C}{B_0}\left(\frac{\Bar{\iota}_0^2}{2}\Delta_0+\Delta_0''\right).
\end{equation*}
A similar consideration applies to the sine component. \par
Demanding periodicity on the pressure tensor rules the general solution of the equation out. This is so, again, because of the generally irrational nature of $\Bar{\iota}_0$, and the fact that the driving term (the RHS of the equation) is periodic. Thus, the solution to the harmonic equation will be the particular solution. \par
As presented in the main text, and given that we are looking for periodic solutions, it is convenient to use a Fourier series for $\Delta_0$ in $\phi$. This way, the second derivative with respect to $\phi$ becomes $-n^2$ and the expressions in (\ref{eq:d11C}) and (\ref{eq:d11S}) follow. \par
This procedure may be continued to higher order. To see this, it is convenient to rewrite Equation I in the form,
\begin{equation}
    \frac{1}{B^2}(\partial_\phi+\Bar{\iota}\partial_\chi)\left(\Delta+\frac{p_\perp}{B^2}\right)=\frac{\Bar{\iota}}{2}\partial_\chi\left(\frac{1}{B^2}\right)\left[\frac{2 p_\perp}{B^2}+\Delta\right]. \label{eq:eqIApp}
\end{equation}
It is clear that, for a given order $n$, the terms containing the largest order functions will have the structure,
\begin{equation*}
    B_0\left(\Delta_n+B_0 p_n\right)'+\Bar{\iota}_0B_0\partial_\chi\left(\Delta_n+B_0 p_n\right)=\dots,
\end{equation*}
where the dots represent combinations of lower order functions. \par
This form of the equation shows that, to arbitrary order, a solution for $\Delta_n$ can be constructed by looking at the particular solution of the corresponding harmonic equation. The special case of $\Delta_{n0}$ is evident from the presented generalisation. \par
For completeness, we present the expansion to order $\epsilon^2$ as well. The non-harmonic term reads,
\begin{align}
    (B_0p_{20}^C&+\Delta_{20}^C)'=-\frac{\Bar{\iota}_0}{4}\frac{B_{11}^C}{B_0}(4B_0p_{11}^S+3\Delta_{11}^S)+\frac{(B_{11}^C)^2\Delta_0'}{2B_0^2}+\nonumber\\
    &+\frac{B_{20}^C}{B_0}\Delta_0'-\frac{B_{11}^C}{2B_0}(2B_0p_{11}^C+\Delta_{11}^C)',
\end{align}
which satisfies the solubility condition without any additional requirement on coefficients. The SHO-like part looks as expected,
\begin{gather}
    \mathcal{A}+2\Bar{\iota}_0(BB_0p_{22}^S+\Delta_{22}^S)+(B_0p_{22}^C+\Delta_{22}^C)'=0\\
    \mathcal{B}-2\Bar{\iota}_0(BB_0p_{22}^C+\Delta_{22}^C)+(B_0p_{22}^S+\Delta_{22}^S)'=0.
\end{gather}
Both $\mathcal{A}$ and $\mathcal{B}$ are a combination of lower order functions of $p$ and $\Delta$,
\begin{gather*}
    \mathcal{A}=\Bar{\iota}_0\left[-\frac{B_{22}^S}{B_0}\Delta_0+\frac{B_{11}^C}{4B_0}(4B_0p_{11}^S+\Delta_{11}^S)\right]-\frac{(B_{11}^C)^2}{2B_0^2}\Delta_0'-\\
    -\frac{B_{22}^C}{B_0}\Delta_0'+\frac{B_{11}^C}{2B_0}(2B_0p_{11}^C+\Delta_{11}^C)'\\
    \mathcal{B}=\Bar{\iota}_0\left[\frac{B_{22}^C}{B_0}\Delta_0-\frac{B_{11}^C}{4B_0}(4B_0p_{11}^C+\Delta_{11}^C)\right]-\frac{B_{22}^S}{B_0}\Delta_0'+\\
    +\frac{B_{11}^C}{2B_0}(2B_0p_{11}^S+\Delta_{11}^S)'.
\end{gather*}

\subsection*{Equation II}
Let us look more carefully at order $\epsilon^n$. To do so in an efficient way, we will rewrite Equation II inspired by the behaviour at lowest orders,
\begin{equation}
    (\partial_\phi+\Bar{\iota}\partial_\chi)\left[B_\theta(1-\Delta)\right]=B_\alpha B^2\left[\frac{1}{B^4}\partial_\chi p_\perp-\frac{1}{2}\Delta\partial_\chi\left(\frac{1}{B^2}\right)\right].
\end{equation}
The LHS of this equation resembles that of (\ref{eq:eqIApp}). It thus follows that the construction of a solution for $B_\theta(1-\Delta)$ will be analogous to that for $B_0p+\Delta$. The particular solution of a SHO equation for the pairs $B_{\theta n m}^{C/S}$ for $0<m<n$ would have to be found, with a simple $\phi$ derivative for $B_{\theta n0}$. In the latter case the equation depends on $p_{n-1}$, rather than $p_n$.  \par
A critical point in Equation II was dropping the largest harmonic constraints (those that originally contained $B_{\theta nn}$). We claimed that these were equivalent to constraints from Equation III, and we shall now prove it. \par
To construct the largest harmonic, the saturation of the Fourier expansions is required, that is, if $p_n$ appears, then it will have to do so as $p_{nn}$. This is so because $\cos k\sin l=[\sin(k+l)-\sin(k-l)]/2$, and thus only if all harmonics are maximized will the largest harmonic be possibly constructed.
\par 
With this in mind, we look at order $\epsilon^n$ of Equation II and $\epsilon^{n-2}$ of Equation III. The relevant saturated, largest harmonic terms are then for Equation II
\begin{equation*}
    -\frac{B_\alpha}{B^2}\partial_\chi p_\perp+\frac{B_\alpha}{2}B^2\Delta\partial_\chi\left(\frac{1}{B^2}\right)=0,
\end{equation*}
where all $B_\theta$ terms have been dropped, as $B_{\theta nn}=0$; and for Equation III,
\begin{equation*}
    -\frac{1}{2}B^2B_\alpha\Delta \partial_\psi\left(\frac{1}{B^2}\right)+\frac{B_\alpha}{B^2}\partial_\psi p_\perp=0.
\end{equation*}
For the latter only those terms involving flux derivatives are allowed. Now, given that we are only concerned with saturated functions, the following relations hold:
\begin{gather*}
    \partial_\psi f_{nn}=\frac{n}{2\epsilon^2}f_{nn} \\
    \partial_\chi f_{nn}=n\Bar{f}_{nn},
\end{gather*}
where as usual, the notation $\Bar{f}_{nn}$ indicates that $\cos n\chi\mapsto-\sin n\chi$ and $\sin n\chi\mapsto\cos n\chi$. It is then clear that, dropping the ordering $\epsilon$ factors, 
\begin{gather*}
    \cos n\chi~ \mathrm{component~of~II} \iff -\frac{1}{2}\sin n\chi ~\mathrm{component~of~III} \\
    \sin n\chi~ \mathrm{component~of~II} \iff \frac{1}{2}\cos n\chi ~\mathrm{component~of~III}. 
\end{gather*}
We have thus shown that the equations are no different, and we may safely drop the two components of II.

\subsection*{Equation III}
Equation III may conveniently be rewritten in the following form:
\begin{align}
    \frac{B_\alpha}{B^2}\partial_\psi p_\perp=(\partial_\phi+\Bar{\iota}\partial_\chi)\left[B_\psi(1-\Delta)\right]+(\Delta-1)(\partial_\psi B_\alpha-\nonumber\\
    -\Bar{\iota}'B_\theta)+\frac{1}{2}B^2B_\alpha\Delta\partial_\psi\left(\frac{1}{B^2}\right). 
\end{align}
Given the flux derivative of the pressure in the LHS, one may easily use Equation III to obtain an expression for $p_n$ to an arbitrarily high order, with all the functions in the RHS being lower order. Also, because of the derivative, the equation to order $\epsilon^{n-2}$ describes $p_n$. The functions that $p_n$ depends on in this construction would be $B_{\psi n-2}, \Delta_{n-1}$ and $B_{\theta n-2}$, but also $p_{n-1}$. For completeness, we write explicitly,
\begin{align}
    B_{\alpha0}B_0\frac{n}{2}p_{n,m}=\left[B_{\psi n-2,m}(1-\Delta_0)\right]'+\nonumber\\
    +\Bar{\iota}_0(1-\Delta_0)m\Bar{B}_{\psi n-2,m}+\dots \label{eq:IIIAnGen}
\end{align}

\section*{Appendix F. Exact cancellation in \textit{looped} equations}
The main purpose of this Appendix is to show that the higher order functions from the \textit{loop} equations cancel exactly when the equations are substituted one into the other. To show this, we shall take $B_{\theta n}$ to be the variable whose solution we seek. To eliminate $p_n$ and $B_{\psi n-2}$ in favour of $B_{\theta n}$ in Equation II, we need to use the closed forms for the former. \par
Start with the equation for $B_{\psi n-2}$; from (\ref{eq:CpBpCBtS}) and (\ref{eq:CpBpSBtC})
\begin{gather*}
    B_{\psi n-2,m}^S=\frac{n}{2m}B_{\theta n,m}^C+\dots \\
    B_{\psi n-2,m}^C=-\frac{n}{2m}B_{\theta n,m}^S+\dots,
\end{gather*}
where the dots depend generally on $X_{n-1},~Y_{n-1},~Z_{n-1},~B_{\psi n-3}$ and $B_{\theta n-1}$. The harmonic components of $p_{nm}$ can be written using (\ref{eq:IIIAnGen}) as,
\begin{align*}
    B_{\alpha0}B_0\frac{n}{2}p_{n,m}^C=\left[B_{\psi n-2,m}^C(1-\Delta_0)\right]'+\\
    +\Bar{\iota}_0(1-\Delta_0)mB_{\psi n-2,m}^S+\dots \\
    B_{\alpha0}B_0\frac{n}{2}p_{n,m}^S=\left[B_{\psi n-2,m}^S(1-\Delta_0)\right]'-\\
    ~~~~~~~~~~-\Bar{\iota}_0(1-\Delta_0)mB_{\psi n-2,m}^C+\dots 
\end{align*}
where the dots represent some combination of $ \Delta_{n-1}$,$~p_{n-1}$, $B_{\theta n-2}$ and lower. Putting these two sets of equations together,
\begin{align*}
    B_{\alpha0}&B_0p_{n,m}^C=\\
    &=-\frac{1}{m}\left[B_{\theta n,m}^S(1-\Delta_0)\right]'+\Bar{\iota}_0(1-\Delta_0)B_{\theta n,m}^C+\dots \\
    B_{\alpha0}B_0&\frac{n}{2}p_{n,m}^S=\\
    &=\frac{1}{m}\left[B_{\theta n,m}^C(1-\Delta_0)\right]'+\Bar{\iota}_0(1-\Delta_0)B_{\theta n,m}^S+\dots 
\end{align*}
which is equivalent to writing,
\begin{equation*}
    B_{\alpha0}B_0p_{n,m}=-\frac{1}{m}\left[\Bar{B}_{\theta n,m}(1-\Delta_0)\right]'+\Bar{\iota}_0(1-\Delta_0)B_{\theta n,m}+\dots 
\end{equation*}
where now the dots represent some combination of $X_{n-1},~Y_{n-1},~Z_{n-1},~\Delta_{n-1},~p_{n-1}$, as well as $B_{\psi n-3}$ and $B_{\theta n-1}$. \par
As a final step, we substitute $p_{nm}$ in the original Equation II. It is evident that all the terms that involve the function $B_{\theta nm}$ exactly cancel! So for each value $m$, there remain two independent equations, in principle, on the lower order functions represented by the dots: $X_{n-1},~Y_{n-1},~Z_{n-1},~\Delta_{n-1}$ and $ ~p_{n-1}$, as well as $B_{\psi n-3}$ and $B_{\theta n-1}$.  \par
The precise cancellation of the order $n$ $B_\theta$ terms occurs systematically at all orders. As a result, the $n$ constraints are to be taken as equations for the functions just mentioned.  \par
The $m=0$ case deserves a slightly different consideration (see for instance the $1/m$ factors in the previous expressions). In fact, when $m=0$ the zeroth harmonic of Equation II is truly an equation for $B_{\theta n+1,0}$, which does not drop out from the equation. This observation is based on the fact that $p$ only appears at an order lower, and thus the substitutions that are required for the other \textit{looped} harmonic equations are not necessary.

\hfill 
\section*{Acknowledgements}
We are grateful to P. Helander, J. Burby, P. Constantin, T. Drivas, D. Ginsberg, N. Kallinikos, and R. MacKay for stimulating discussions This research is primarily supported by a grant from the Simons Foundation/SFARI (560651, AB). 

\section*{Data availability}
Data sharing is not applicable to this article as no new data were created or analyzed in this
study.

\bibliography{avoidOverGB}

\end{document}